\documentclass[preprint,letter,amssymb]{revtex4}
\usepackage{graphicx,bm,amsmath,epsfig}

\begin{document}


\def\reff#1{(\ref{#1})}
\newcommand{\be}{\begin{equation}}
\newcommand{\ee}{\end{equation}}
\newcommand{\<}{\langle}
\renewcommand{\>}{\rangle}

\def\spose#1{\hbox to 0pt{#1\hss}}
\def\ltapprox{\mathrel{\spose{\lower 3pt\hbox{$\mathchar"218$}}
 \raise 2.0pt\hbox{$\mathchar"13C$}}}
\def\gtapprox{\mathrel{\spose{\lower 3pt\hbox{$\mathchar"218$}}
 \raise 2.0pt\hbox{$\mathchar"13E$}}}

\def\bsigma{\mbox{\protect\boldmath $\sigma$}}
\def\bpi{\mbox{\protect\boldmath $\pi$}}
\def\smfrac#1#2{{\textstyle\frac{#1}{#2}}}
\def\smhalf{ {\smfrac{1}{2}} }

\newcommand{\re}{\mathop{\rm Re}\nolimits}
\newcommand{\im}{\mathop{\rm Im}\nolimits}
\newcommand{\tr}{\mathop{\rm tr}\nolimits}
\newcommand{\fr}{\frac}

\def\Z{{\mathbb Z}}
\def\R{{\mathbb R}}
\def\C{{\mathbb C}}

\title{Fluid-fluid demixing curves for colloid-polymer mixtures \\
in a random colloidal matrix}

 \author{Mario Alberto Annunziata}
 \address{CNR, Istituto dei Sistemi Complessi \\
         (Area della Ricerca di Roma Tor Vergata) \\
          Via del Fosso del Cavaliere 100, I-00133 Roma, Italy \\
          and INFN, Sezione di Pisa \\
          L.go Pontecorvo 3, I-56127 Pisa, Italy \\
   e-mail: {\tt m.annunziata@isc.cnr.it}}
 \author{ Andrea Pelissetto }
 \address{
   Dipartimento di Fisica, Universit\`a degli Studi di Roma ``La Sapienza'' \\
   and INFN -- Sezione di Roma I  \\
   Piazzale A. Moro 2, I-00185 Roma, Italy \\
  e-mail: {\tt Andrea.Pelissetto@roma1.infn.it }}


\begin{abstract}

We study fluid-fluid phase separation in a colloid-polymer mixture
adsorbed in a colloidal porous matrix close to the $\theta$ point. For this
purpose we consider the Asakura-Oosawa model in the presence of 
a quenched matrix of colloidal hard spheres. We study the dependence of 
the demixing curve on the parameters that characterize
the quenched matrix, fixing the polymer-to-colloid size ratio 
to 0.8. We find that, to a large extent, demixing curves 
depend only on a single parameter $f$, 
which represents the volume fraction which 
is unavailable to the colloids. We perform Monte Carlo simulations 
for volume fractions $f$ equal to 40\% and 70\%, finding that the 
binodal curves in the polymer and colloid packing-fraction plane 
have a small dependence on disorder. The critical point instead
changes significantly: for instance, the 
colloid packing fraction at criticality increases
with increasing $f$. Finally, we observe for some values of the parameters
capillary condensation of the colloids: a bulk colloid-poor phase is 
in chemical equilibrium with a colloid-rich phase in the matrix.
\bigskip 

PACS: 61.25.Hq, 82.35.Lr


\end{abstract}

\maketitle
\thispagestyle{empty}   

\clearpage

\section{Introduction}

The study of the fluid phases in mixtures of colloids and 
nonadsorbing neutral polymers has become increasingly 
important in recent years; see 
Refs.~\cite{Poon-02,FS-02,TRK-03,MvDE-07,FT-08,ME-09} for recent reviews,
Refs.~\cite{LLPR-03,VvDV-03,SCSZ-03,KSMH-04,LCP-04,HEAD-05,KSH-05,
ZvD-06,LPKCSBFE-09,MvDEGH-09-10} for experiments, and 
Refs.~\cite{GHR-83,LPPSW-92,MF-94,Sear-97-02,DBE-99,FS-00,SLBE-00,
BLH-02,DvR-02, DLL-02,BLM-03,MJLBR-03,Tuinier-03,PVJ-03,VH-04bis,
VHB-05,VS-05,VJDL-05,
Bryk-05,FS-05,PH-05,CVPR-06,FT-07,ZVBHV-09} for theoretical investigations.
These systems show a very interesting phenomenology, which only depends 
to a large extent on the nature of the solvent and 
on the ratio $q \equiv R_g/R_c$, where $R_g$ is the radius of 
gyration of the polymer and $R_c$ is the radius of the colloid.
Experiments and numerical simulations indicate that polymer-colloid mixtures
have a solid colloidal phase for large enough colloidal concentrations
and a corresponding fluid-solid coexistence. 
Much less obvious is the presence of a fluid-fluid coexistence of a 
colloid-rich, polymer-poor phase (colloid liquid) with a 
colloid-poor, polymer-rich phase (colloid gas). Extensive theoretical and
experimental work
indicates that such a transition occurs only if the size of the polymers is 
sufficiently large, i.e. for $q > q^*$, where
\cite{Poon-02, DLL-02,CVPR-06} $q^* \approx 0.3$-0.4. 

At least
qualitatively, many aspects of the behavior of colloid-polymer suspensions
can be understood by using the Asakura-Oosawa (AO) model
\cite{AO-54,Vrij-76}, which gives a coarse-grained description of the 
mixture. The polymers are treated as an ideal gas 
of point particles of radius $R_p$ (which is usually identified with the 
radius of gyration) which interact with the colloids by means 
of a simple hard-core potential. 
This model is extremely crude since it ignores the polymeric structure and 
polymer-polymer repulsion, which is relevant in the good-solvent regime.
Nonetheless, it correctly predicts
polymer-colloid demixing as a result of the entropy-driven 
effective attraction (depletion interaction) 
between colloidal pairs due to the presence 
of the polymers 
\cite{MF-94,DBE-99,VH-04bis,VHB-05,BLH-02,DvR-02,GHR-83,LPPSW-92,SLBE-00}. 
It is not, however, quantitatively predictive for polymers in the 
good-solvent regime.
For instance, 
at a given colloid packing fraction, the AO model predicts the binodal curve
to be at a polymer volume fraction which is significantly lower than that 
observed experimentally.
In order to reproduce the experimental results one can use 
realistic atomistic models for the polymers, but this is a very difficult 
task from a numerical point of view. In the colloid regime 
$q \lesssim 1$, it is much easier, and still 
provides good results, to use
coarse-grained models in which polymers are modelled 
as point particles (as in the AO model) interacting with repulsive 
soft pair potentials 
\cite{DLL-02,BLH-02,VS-05,ZVBHV-09}, which have either 
a phenomenological origin or are derived by means of exact
coarse-graining procedures. Nonetheless, at least for $q\lesssim 1$
(colloidal regime), the AO model is expected to provide 
quantitatively correct results for colloid-polymer 
solutions close to the $\theta$ point.  Indeed, in this regime 
polymers show an approximate ideal behavior and can be described 
quite reasonably as noninteracting random walks, as does the AO model
\cite{footnote-theta}. 

In this paper we wish to study the demixing of colloid-polymer mixtures
in porous materials, which are characterized by a highly
interconnected porous structure. They have important technological applications,
for instance in catalysis and gas separation and purification
\cite{footnote-expt-porousmat}. 
Examples are the Vycor glasses,
in which pore sizes range from 1 nm to 100 nm, and high-porosity 
systems like silica gels 
(xerogels and aerogels), which are produced by means of silica
sol-gel processes. AO colloid-polymer mixtures in a porous matrix have been
studied in Refs.~\cite{SSKK-02,VBL-06,VBL-08,PVCL-08,Vink-09} by means 
of density-functional theory, integral equations, and 
Monte Carlo (MC) simulations. The nature of the 
critical transition has been fully clarified 
\cite{VBL-06,VBL-08,PVCL-08,Vink-09}:
if obstacles are random and there is a preferred affinity 
of the quenched obstacles to one of the phases,
the transition is in the same universality class as that 
occurring in the random-field Ising model, 
in agreement with a general argument by de Gennes \cite{deGennes-84}.
If these conditions are not satisfied, standard Ising or randomly
dilute Ising behavior is observed instead, see Refs.~\cite{DSLP-08,FV-11}.
On the other hand, little is known on how demixing is 
influenced by the amount of disorder and by its nature
(for a polymer matrix some results for the critical-point behavior 
as a function of the amount of disorder are reported in Ref.~\cite{VBL-06}).
In this paper 
porosity is introduced by considering a quenched matrix of 
hard spheres of radius $R_{\rm dis}$. We will compute the binodal curves in 
terms of the polymer and colloid packing fractions for different 
ratios $R_{\rm dis}/R_c$ and for different disorder concentrations
with the purpose of determining how these parameters affect the 
location of the demixing transition and of the critical 
(second-order) transition point.
We will not instead perform a detailed study of the $q$ dependence
and we shall set $q = 0.8$ as in Ref.~\cite{PVCL-08}. 
This work complements the results of Ref.~\cite{SSKK-02}, which instead
studied the $q$ dependence for a single value of $R_{\rm dis}/R_c$,
$R_{\rm dis}/R_c = 1$, and of the disorder concentration. 

The paper is organized as follows. In Sec.~\ref{sec2} we discuss the model and
the relevant variables. In Sec.~\ref{sec3} we present our numerical results.
Our conclusions are presented in Sec.~\ref{sec4}. In App.~\ref{app} we 
present some details on the MC calculation.

\section{The model} \label{sec2}

In the AO model polymers and colloids are modelled as spheres of radii $R_p$ and
$R_c$, respectively. We assume hard-sphere interactions between colloid and 
colloid-polymer pairs; the pair potentials are given by
\begin{eqnarray}
u_{cc}(r) &=& \begin{cases}
   \infty & \text{$\qquad$ for $r < 2 R_c$,} \cr 
   0 & \text{$\qquad$ for $r \ge 2 R_c$,}
   \end{cases} \nonumber \\ [3mm]
u_{cp}(r) &=& \begin{cases}
   \infty & \text{$\qquad$ for $r < R_c + R_p$,} \cr 
   0 & \text{$\qquad$ for $r \ge R_c + R_p$,}
   \end{cases} \nonumber \\ [3mm]
u_{pp}(r) &=&  0,
\end{eqnarray}
where $r$ is the center-to-center distance.
We consider a cubic box of size $L$ and we characterize the thermodynamic 
phases in terms of the packing fractions 
\begin{equation}
\eta_p \equiv {4 \pi R_p^3 N_p\over 3 L^3} \qquad\qquad
\eta_c \equiv {4 \pi R_c^3 N_c\over 3 L^3},
\end{equation}
where $N_p$ and $N_c$ indicate the number of polymers and of colloids in the 
box, respectively. 

The phase behavior of the AO model
has been extensively studied. It strongly depends on the polymer-to-colloid
size ratio $q \equiv R_p/R_c$. For small values of $q$ 
the demixing transition is unstable and only the fluid-solid transition occurs.
Fluid-fluid demixing occurs \cite{Poon-02, DLL-02,CVPR-06} 
for $q\gtrsim 0.3$-0.4.
In this work we have not investigated 
the $q$ dependence of the binodal curve, since our main objective is the 
analysis of the role of quenched disorder. We have thus fixed $q = 0.8$, 
as in Ref.~\cite{PVCL-08},
at the boundary between the colloid and the protein regimes.

Disorder has been introduced by considering a colloidal quenched matrix
which has a hard-sphere interaction both with the colloids and the polymers.
In practice, we choose a disorder concentration $c_{\rm dis}$ and 
randomly distribute $N_{\rm dis} = c_{\rm dis} L^3$ 
nonoverlapping spheres of radius $R_{\rm dis}$ in the box. 
The position of these 
spheres is assumed to be fixed (quenched). Colloids and polymers can only
move outside the quenched matrix, which means that the spheres belonging
to the matrix and the freely moving particles interact with 
pair potentials
\begin{eqnarray}
u_{c,\rm dis}(r) &=& \begin{cases}
   \infty & \text{$\qquad$ for $r < R_c + R_{\rm dis}$,} \cr 
   0 & \text{$\qquad$ for $r \ge R_c + R_{\rm dis}$,}
   \end{cases} \nonumber \\ [3mm]
u_{p,\rm dis}(r) &=& \begin{cases}
   \infty & \text{$\qquad$ for $r < R_p + R_{\rm dis}$,} \cr 
   0 & \text{$\qquad$ for $r \ge R_p + R_{\rm dis}$.}
   \end{cases} 
\label{pair-pot-disorder}
\end{eqnarray}
Note that the matrices considered here are different from 
those discussed in Refs.~\cite{VBL-06,VBL-08}. The main difference is 
that here the matrix consists in hard spheres that cannot intersect 
each other (we name it colloidal matrix). 
On the other hand, in Refs.~\cite{VBL-06,VBL-08} 
the matrix spheres are soft and can freely overlap, as if they were an ideal
gas (hence the name polymer matrix). Second, in those works, 
for a given choice of ${c}_{\rm dis}$, the number $N_{\rm dis}$
is not fixed, but obtained from a Poissonian distribution with 
mean value ${c}_{\rm dis} L^3$. This second difference should not be important
in the infinite-volume limit, since it entails density fluctuations of 
order $1/L^{3/2}$, which vanish as $L\to\infty$. 

In the simple model we consider, disorder is characterized by two parameters,
the reduced concentration 
$\hat{c} \equiv c_{\rm dis} R_c^3$ and the ratio $R_{\rm dis}/R_c$. 
However, $\hat{c}$ does not directly characterize the free space available to
the colloids and to the polymers. 
We shall use instead the 
effective matrix filled-space ratio $f$,
which is defined as follows. Consider the region ${\cal R}$
in which the (centers of the) colloids are allowed:
\begin{equation}
{\cal R} = \{ {\bf r}: |{\bf r} - {\bf r}_i| \ge R_c + R_{\rm dis} ,
\hbox{
for all $1\le i\le N_{\rm dis}$ }
\},
\end{equation}
where ${\bf r}_i$ is the position of the $i$-th hard sphere belonging
to the matrix. If $V_{\cal R}$ is the volume of the region ${\cal R}$, 
we define
\begin{equation}
   f \equiv 1 - {[V_{\cal R}]\over L^3},
\end{equation}
where $[V_{\cal R}]$ is the average of $V_{\cal R}$ over the different matrix 
realizations. Note that, for large values of $L$, $[V_{\cal R}]$ is essentially
independent of the matrix realization, a property known as self-averaging.
The parameter $f$ represents the volume fraction that is unavailable to the 
colloids due to the presence of the random matrix and 
can easily be determined by computing the probability of 
inserting a colloid in the otherwise empty matrix. In a completely 
analogous way we can define $f_{\rm pol}$, which characterizes 
the volume fraction unavailable to polymers. Of course, $f > f_{\rm pol}$
in the colloid regime in which $q < 1$, while $f < f_{\rm pol}$ in 
the opposite, protein regime. 

\begin{figure}
\begin{tabular}{cc}
\epsfig{file=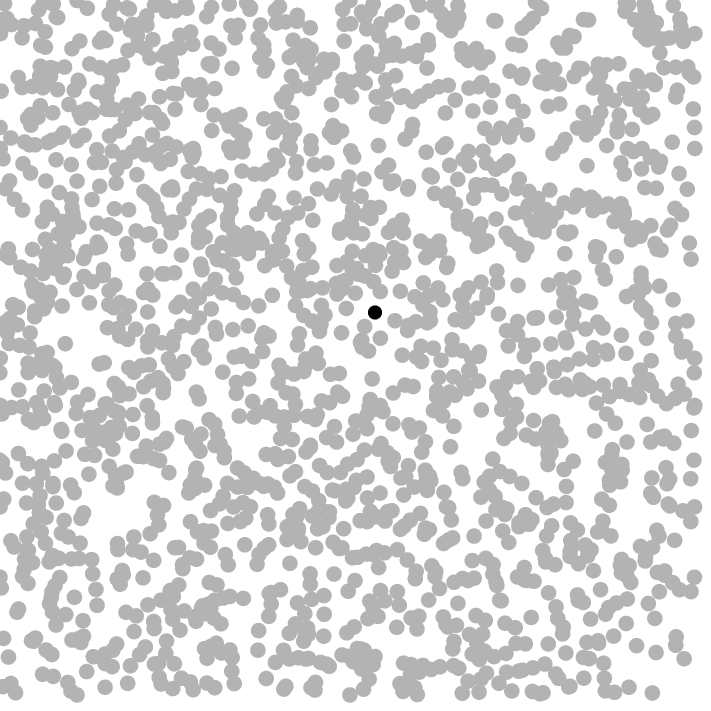,angle=-90,width=7truecm} \hspace{0.5truecm} &
\epsfig{file=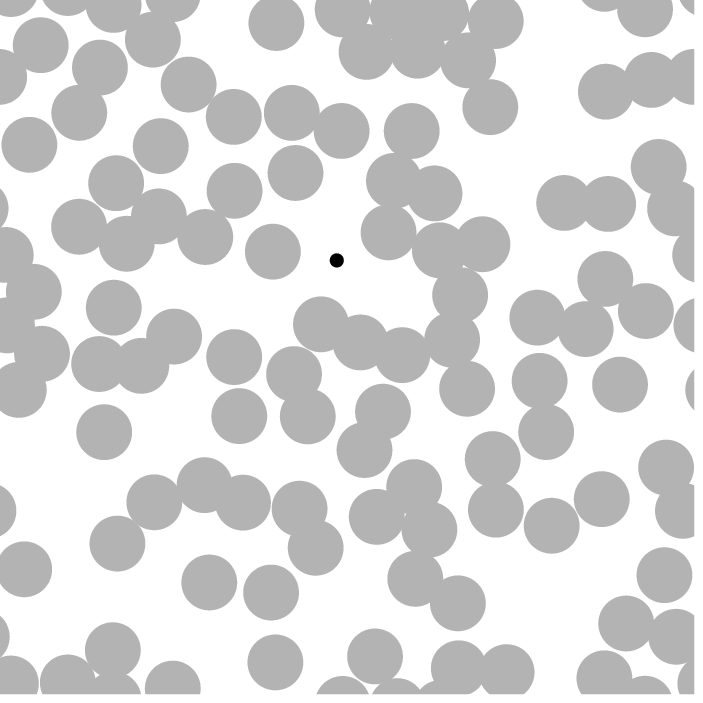,angle=-90,width=7truecm} \\
\end{tabular}
\vspace{1cm}
\caption{Two-dimensional systems with $L/R_c = 50$ and 
 $f = 0.5$. On the left we take 
 $R_{\rm dis}/R_c = 0.1$, while on the right $R_{\rm dis}/R_c = 3.0$.
 The disorder packing fractions $\pi c R_{\rm dis}^2$ 
 are 0.0057 and 0.292 in the two cases, respectively.
 The gray circles of radius
$R_c + R_{\rm dis}$ correspond to the colloid-excluded region
(depletion region)
around each sphere of the quenched matrix:
the centers of the colloids can only belong to the white region. 
We also draw a single colloid (black) 
to show the length scale. 
}
\label{fig-topology-f}
\end{figure}

It is interesting to understand qualitatively how the disorder distribution
changes with $R_{\rm dis}$ at fixed $f$. In Fig.~\ref{fig-topology-f}
we show the matrix for $f = 0.5$ and two values of $R_{\rm dis}$, 
$R_{\rm dis}/R_c = 0.1$ and $R_{\rm dis}/R_c = 3$. To make the figure more 
clear, we consider a two-dimensional system, that is 
a matrix of nonoverlapping disks on a square of area $L^2$.
It is evident that the topology of the matrix is quite different. 
For large $R_{\rm dis}/R_c$ 
the free volume available to the colloids consists in large empty regions 
connected by narrow channels. This is the case of a porous material 
with big interconnected pores. On the other hand, for $R_{\rm dis}/R_c$
small, pores are significantly smaller and
the topology of the network is more complex. 

In order to have demixing, the parameter $f$ 
cannot be arbitrarily close to 1, but should satisfy $f < f^*$, where 
$f^*$ is related to the percolation threshold of the region 
${\cal R}$ in which colloids can move. For $f > f^*$ 
the space ${\cal R}$
divides in disconnected finite regions and thus no phase transition 
is possible. The exact value of $f^*$ is unknown. However, the arguments of 
Ref.~\cite{SZ-70} suggest 
\begin{eqnarray}
    f^* \approx 0.85.
\end{eqnarray}
For the same reasons --- polymers should be able to move in the 
whole space --- the polymer parameter $f_{\rm pol}$ must satisfy
$f_{\rm pol} < f^*$ in order to observe coexistence.

In this paper we shall perform simulations for two values of $f$, 
$f = 0.40$ and $f = 0.70$, the latter being quite close to the 
threshold $f^*$, and for $q = 0.8$, so that 
$f_{\rm pol} < f$. In Table \ref{tab-f} we report the 
reduced concentration $\hat{c}$ and the disorder packing fraction 
$\eta_{\rm dis} \equiv 4 \pi R_{\rm dis}^3 c/3$ for several values of 
$R_{\rm dis}/R_c$. First, we observe that $\hat{c}$ converges to a 
finite positive constant as $R_{\rm dis}/R_c\to 0$. This result 
is quite easy to understand.
If $R_{\rm dis}/R_c \ll 1$, the pair potentials 
(\ref{pair-pot-disorder}) become essentially independent of $R_{\rm dis}$.
Hence, the density becomes essentially independent of $R_{\rm dis}$ 
for $R_{\rm dis}$ small. In the opposite limit
$R_{\rm dis}/R_c \gg 1$, the potentials become essentially 
independent of $R_c$. Hence, in this limit $f$ converges to 
the disorder packing fraction
$\eta_{\rm dis} \equiv 4 \pi R_{\rm dis}^3 c_{\rm dis}/3$.
For instance, for $\eta_{\rm dis} = 0.30$, 
we obtain $f = 0.51,0.40,0.32$ for $R_{\rm dis}/R_c = 5,10,50$. 
Since a liquid hard-sphere phase 
exists only up \cite{HR-67} 
to $\eta \approx 0.49$, for large $R_{\rm dis}/R_c$,
the matrix may belong to different hard-sphere phases, while
still satisfying the condition $f < f^*$.

\begin{table}
\caption{Estimates of the reduced concentration
 $\hat{c}\equiv c_{\rm dis} R_c^3$ and of the 
disorder packing fraction 
$\eta_{\rm dis} \equiv 4 \pi R_{\rm dis}^3 c_{\rm dis}/3$ 
for two values of $f$ and several values of $R_{\rm dis}/R_c$.
}
\label{tab-f}
\begin{tabular}{clclc}
\hline\hline
& \multicolumn{2}{c}{$f = 0.40$} 
& \multicolumn{2}{c}{$f = 0.70$} \\
\multicolumn{1}{c}{$R_{\rm dis}/R_c$} &  
\multicolumn{1}{c}{$\hat{c}$} & 
\multicolumn{1}{c}{$\eta_{\rm dis}$} & 
\multicolumn{1}{c}{$\hphantom{???}$$\hat{c}$} & 
\multicolumn{1}{c}{$\eta_{\rm dis}$} \\
\hline
0.005  &  0.120  & $6.3\cdot 10^{-8}$  & $\hphantom{???}$
 0.283  & $1.5\cdot 10^{-7}$  \\
0.01   &  0.118  & $4.9\cdot 10^{-7}$  & $\hphantom{???}$
 0.279  & $1.2\cdot 10^{-6}$  \\
0.02   &  0.115  & $3.9\cdot 10^{-6}$  & $\hphantom{???}$
 0.271  & $9.1\cdot 10^{-6}$  \\
0.05   &  0.105  & $5.5\cdot 10^{-5}$  & $\hphantom{???}$
 0.248  & $1.3\cdot 10^{-4}$  \\
0.1    &  0.0915  & $3.8\cdot 10^{-4}$  & $\hphantom{???}$
 0.215  & $9.0\cdot 10^{-4}$  \\
0.2    &  0.0700  & $2.3\cdot 10^{-3}$  & $\hphantom{???}$
 0.164  & $5.5\cdot 10^{-3}$  \\
0.4    &  0.0431 & 0.0116 & $\hphantom{???}$ 0.0972 & 0.026  \\
0.6    &  0.0280 & 0.0254 & $\hphantom{???}$ 0.0609 & 0.055 \\
1.0    &  0.0136 & 0.057 & $\hphantom{???}$ 0.0278 & 0.116 \\
\hline\hline
\end{tabular}
\end{table}

\section{Results} \label{sec3}

\subsection{Monte Carlo simulation} \label{sec3.1}

In this work we investigate the effect of disorder on the 
fluid-fluid binodals for $q = 0.8$.  We perform
simulations in the absence of the porous matrix --- our results are 
consistent with those of Refs.~\cite{VHB-05,VS-05} --- and for two values of 
$f$, $f = 0.4$ and $f = 0.7$ 
[note that $\hat{c}(f=0.7)\approx 2 \hat{c}(f=0.4)$],
in cubic boxes $L^3$ with 
$L/R_c = 16$ and 20. In order to obtain quenched averages we consider 
200-400 matrix realizations for each $f$ and $R_{\rm dis}$.

For each value of $f$ we consider a few values of $R_{\rm dis}/R_c$. 
For $f = 0.4$ we present results for $R_{\rm dis}/R_c = 0.2,0.6,1.0$
($f_{\rm pol} = 0.26$, 0.29, and 0.31, respectively),
while for $f=0.7$, we use $R_{\rm dis}/R_c =0.2$ and 1.0
($f_{\rm pol} = 0.50$ and 0.57, respectively). It should be noted 
that we are limited by computer power in further decreasing or increasing
the ratio  $R_{\rm dis}/R_c$. Indeed, if we further decrease the ratio,
the disorder density increases, see Table~\ref{tab-f}, and so does the number 
of matrix particles and the computational work. On other hand, if we increase
$R_{\rm dis}/R_c$ beyond 1, we should consider quite large systems in order
to avoid large size effects, which is unfeasible with our present computer
power. 

In order to determine the coexistence curves we perform a grand-canonical 
simulation. The grand partition sum for each disorder realization is
\begin{equation}
\Xi(V,z_p,z_c) = \sum_{N_p,N_c} z_p^{N_p} z_c^{N_c} Q(V,N_p,N_c),
\label{def-Xi}
\end{equation}
where $Q(V,N_p,N_c)$ is the configurational partition function of 
a system of $N_p$ polymers and $N_c$ colloids in a volume $V$,
and $z_p$ and $z_c$ are the corresponding fugacities. 
In Eq.~(\ref{def-Xi}) we normalize $Q(V,N_p,N_c)$ so that 
$Q(V,1,0) = Q(V,0,1) = V$, hence $z_p$ and $z_c$ are dimensionful
parameters. We quote our results in terms of the dimensionless 
combinations $z_c R^3_c$ and 
\begin{equation}
\eta_p^r = {4\pi\over 3} z_p R_p^3.
\end{equation} 
The quantity $\eta_p^r$ represents the polymer reservoir packing fraction.

In the presence of a first-order transition, standard local algorithms
are unable to sample correctly both phases in the simulation. We 
therefore combine the grand-canonical algorithm with 
the umbrella sampling and
the simulated-tempering method \cite{TV-77,MP-92},  
as discussed in App.~\ref{app}. 
Insertions and deletions of colloids and polymers are performed 
by using  the cluster 
moves introduced by Vink and Horbach \cite{VH-04bis,Vink-04}.

\subsection{Quenched coexistence curve}
\label{sec3.2}

The main purpose of this work is the determination of the 
disorder-averaged coexistence curve. In order to define it precisely, 
let us define the disorder-averaged colloid and polymer numbers 
\begin{eqnarray}
N_{c,\rm av}(V,z_p,z_c) = z_c {\partial \over \partial z_c} 
   \left[ \ln \Xi(V,z_p,z_c)\right], \nonumber \\
N_{p,\rm av}(V,z_p,z_c) = z_p {\partial \over \partial z_p} 
   \left[ \ln \Xi(V,z_p,z_c)\right], 
\end{eqnarray}
where $[\cdot ]$ indicates the average over the matrix realizations. 
In the presence of first-order transitions, there is a line $z_c = z_c^*(z_p)$
in the $(z_p,z_c)$ plane where these two functions become discontinuous
in the infinite-volume limit. 
In other words, for $z_p > z_{p,\rm crit}$ we have 
\begin{eqnarray}
\lim_{\epsilon\to 0+} \lim_{V\to\infty} 
    N_{c,\rm av}(V,z_p,z_c^*(z_p)+ \epsilon)/V = 
               c_{c,\rm liq}, \nonumber \\
\lim_{\epsilon\to 0+} \lim_{V\to\infty} 
       N_{p,\rm av}(V,z_p,z_c^*(z_p)+ \epsilon)/V = 
               c_{p,\rm liq}, \nonumber \\
\lim_{\epsilon\to 0+} \lim_{V\to\infty} 
       N_{c,\rm av}(V,z_p,z_c^*(z_p)- \epsilon)/V = 
               c_{c,\rm gas}, \nonumber \\
\lim_{\epsilon\to 0+} \lim_{V\to\infty} 
       N_{p,\rm av}(V,z_p,z_c^*(z_p)- \epsilon)/V = 
               c_{p,\rm gas}. 
\end{eqnarray}
The pair $c_{p,\rm liq}$, $c_{c,\rm liq}$ gives the polymer and 
colloid concentrations in the colloid-liquid phase at coexistence, while 
$c_{p,\rm gas}$ and $c_{c,\rm gas}$ correspond to the 
colloid-gas polymer-rich phase. 

In the MC simulations the position of the demixing curve
can be determined by studying the disorder averaged histograms of $N_c$
and $N_p$, which are defined as 
\begin{eqnarray}
h_{c,\rm ave} (N_{c,0},z_p,z_c) &\equiv & 
  \left[ \left\< \delta(N_c,N_{c,0})\right\>_{GC,z_p,z_c}\right] ,
\\
h_{p,\rm ave} (N_{p,0},z_p,z_c) &\equiv & 
  \left[ \left\< \delta(N_p,N_{p,0})\right\>_{GC,z_p,z_c}\right] ,
\end{eqnarray}
where $\delta({x,y})$ is the Kronecker's delta [$\delta({x,x}) = 1$, 
$\delta({x,y}) = 0$ for $x\not= y$] and $\left\<\cdot \right\>_{GC,z_p,z_c}$
is the grand-canonical ensemble average. In the two-phase region 
the histograms show a double-peak structure. In order to obtain 
$z_c^*$ at fixed $z_p$ in a finite volume, several different methods can
be used. We followed two different recipes, the equal-area and the 
equal-height methods.  In the first case, we define $z_c^*$ as 
the value of the colloid fugacity at which
the area below the two peaks is equal. For instance, if we consider
the colloid-number distribution, we first compute the 
position $N_{\rm min}$ of the minimum between the two peaks and then 
require $z_c^*$ to be the value of the colloid fugacity at which 
\be
\sum_{N_c < N_{\rm min}} h_{c,\rm ave} (N_c,z_p,z_c^*) = 
\sum_{N_c > N_{\rm min}} h_{c,\rm ave} (N_c,z_p,z_c^*).
\label{equal-area}
\ee
Equivalently, one can use the polymer distribution 
$h_{p,\rm ave} (N_c,z_p,z_c)$.
In the second method we identify $z_c^*$ as the value of the fugacity 
at which the two peaks have the same height. Once $z_c^*$  has been 
obtained, the colloid and polymer number at the transition are defined as 
the positions of the maxima of the histograms. 

\begin{figure}
\epsfig{file=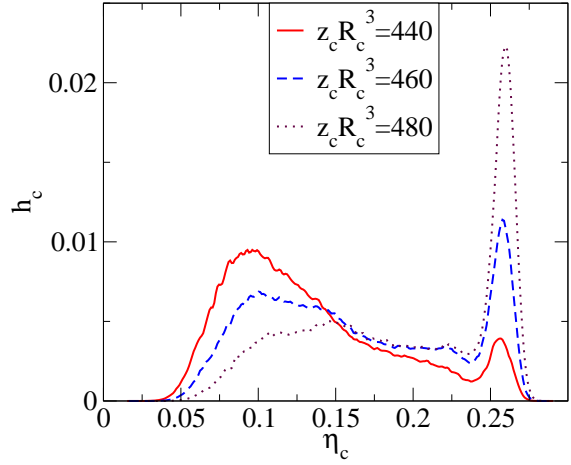,angle=0,width=8truecm} 
\caption{Colloid histogram $h_{c,\rm ave}$ for 
  $L/R_c=16$, $f = 0.7$, 
  $R_{\rm dis}/R_c = 1.0$, $\eta_p^r \approx 1.82$ ($z_p R_c^3 = 0.85$), 
  and some values of $z_c$ close to $z_c^*$. 
}
\label{critical-fugacity2}
\end{figure}

Since we have two different histograms to analyze, we obtain two 
different estimates of the colloid fugacity at coexistence: 
an estimate $z_c^*(c)$ is obtained from the analysis of the 
colloid-number histograms, while  $z_c^*(p)$ is obtained from the analysis
of the polymer-number histograms. For $R_{\rm dis}/R_c=0.2,0.6$
the two estimates are quite close 
and provide consistent estimates of the colloid and polymer 
packing fractions at coexistence, although
the equal-area method is more thermodynamically
consistent. Indeed,
the differences $|z_c^*(c) - z_c^*(p)|$ computed with 
the equal-area method are always smaller than those computed with the 
second prescription. For $R_{\rm dis}/R_c=1.0$, we have been unable to 
apply the area method. The difficulties can be understood by
looking at Fig.~\ref{critical-fugacity2}, where we report
the colloid histograms for $f = 0.7$ and for the largest value of $z_p$
we consider, $\eta_p^r \approx 1.82$ ($z_p R_c^3= 0.85$). 
While the colloid-liquid peak is quite narrow, 
the colloid-gas peak is very broad and therefore
condition (\ref{equal-area}) is satisfied only when the 
colloid-gas peak is barely visible. However, in this case the definition
of $N_{\rm min}$ is ambiguous and thus $z_c^*$ is determined with 
large uncertainty.  In some cases, it is even impossible
to satisfy the equal-area condition.  Thus, for $R_{\rm dis}/R_c=1.0$
we only use the equal-height method. 
Note that the two methods should give identical results in the infinite-volume 
limit. Hence, the difficulties we observe indicate that for this value
of the parameters finite-size effects are important. The analyses reported in
the following sections confirm these findings. 
It is interesting to note that, at variance with what happens in the bulk,
in the presence of randomness
the order parameter distribution shows two well-separated 
nonoverlapping peaks even at the critical point 
(see Refs.~\cite{VBL-06,VBL-08} for a discussion in
the present context). Thus, it is also possible that the 
difficulties we observe for some values of the parameters 
are related to the fact that they belong to the one-phase region,
even if the finite-size colloid and polymer histograms are bimodal.

\begin{figure}
\begin{tabular}{ll}
\epsfig{file=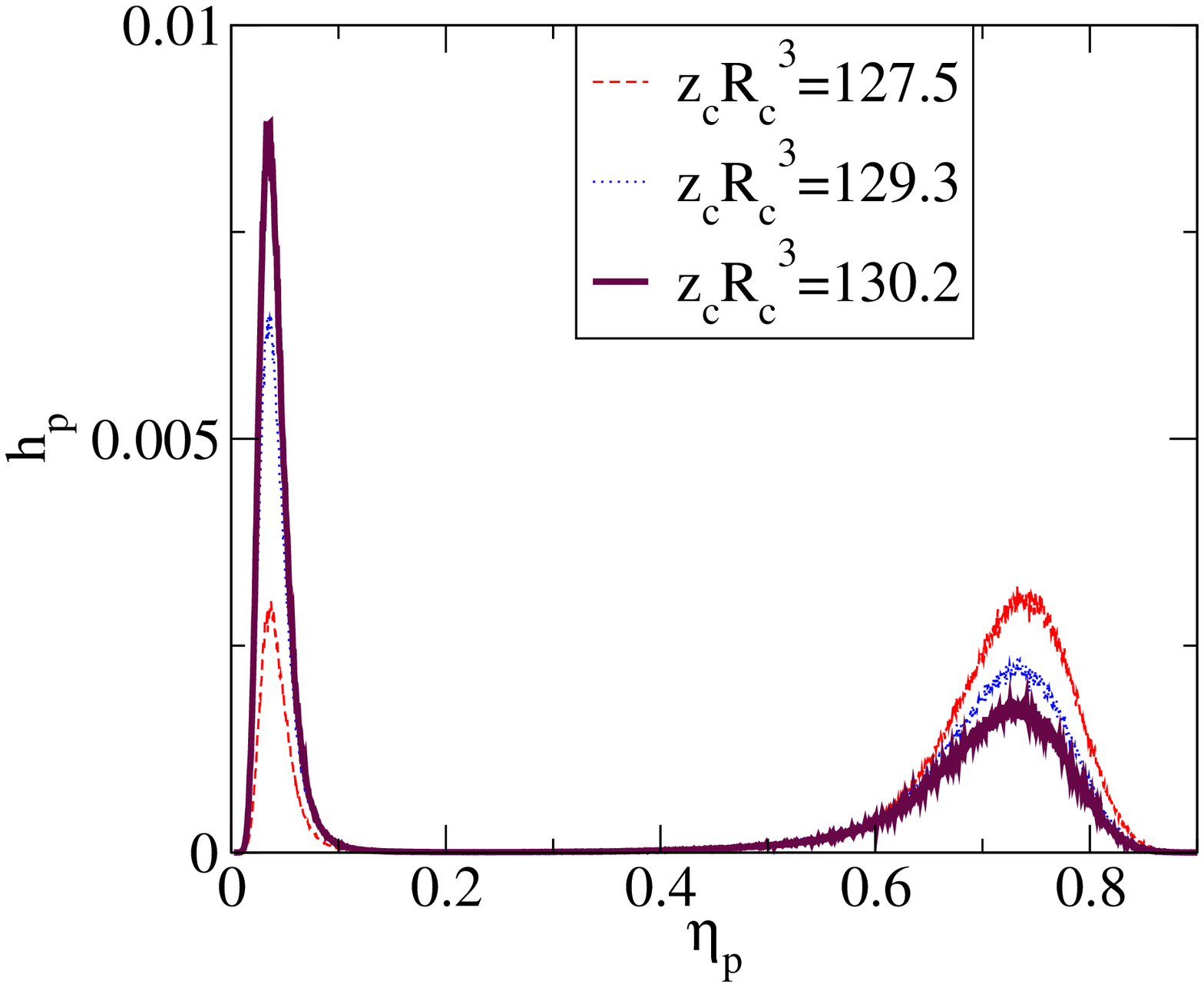,angle=0,width=8truecm} 
  \hspace{-0.0truecm} &
\epsfig{file=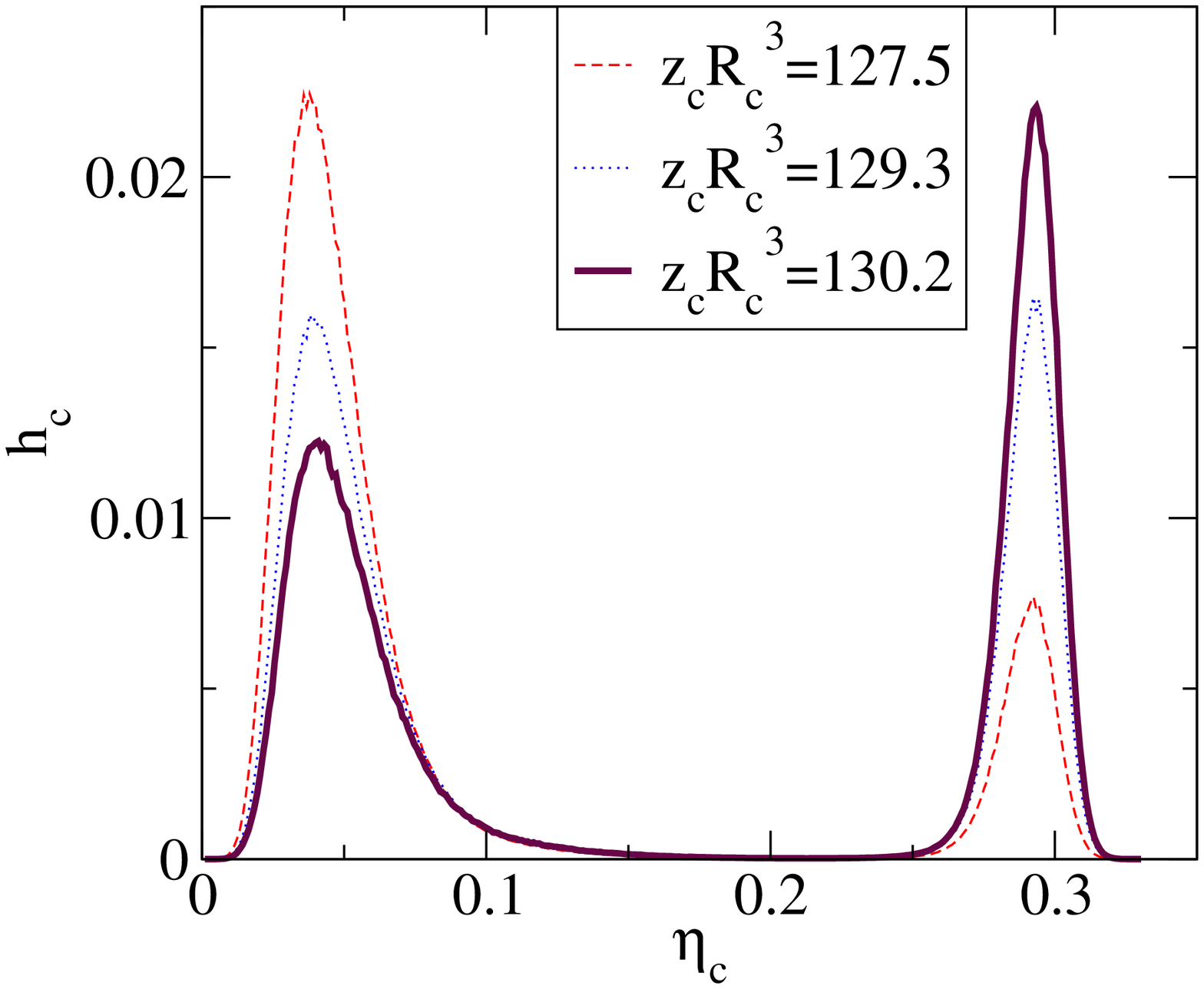,angle=0,width=8truecm} \\
\vspace{-0.2truecm} \\
\end{tabular}
\caption{Polymer histogram $h_{p,\rm ave}$ and 
  colloid histogram $h_{c,\rm ave}$ for $L=16R_c$, $f = 0.4$, 
  $R_{\rm dis}/R_c = 0.6$, $\eta_p^r \approx 1.24$ ($z_p R_c^3 = 0.58$), 
  and several fugacities $z_c R_c^3$.
  The thicker curve corresponds to the coexistence fugacity
  obtained by using the equal-area prescription ($z_c^*R_c^3 = 130.2$), 
  while the other two curves correspond to the estimates
  $z_c^*(c) R_c^3 = 129.3$ and $z_c^*(p) R_c^3 = 127.5$ obtained by using the 
  equal-height method.
}
\label{critical-fugacity}
\end{figure}

To give an idea of the performance of the two methods,
we report the results for $f = 0.4$, $R_{\rm dis}/R_c = 0.6$,
$\eta_p^r \approx 1.24$
($z_p R_c^3= 0.58$), and $L/R_c = 16$, see Fig.~\ref{critical-fugacity}. 
The equal-area method gives 
\begin{equation}
z_c^* R_c^3 \approx 130.2 \qquad \eta_{c,\rm gas} \approx 0.041 \qquad 
                    \eta_{c,\rm liq} \approx 0.294,
\label{coexdata-1}
\end{equation}
from the analysis of the colloid distribution. The analysis of the polymer 
distribution gives the same estimate of $z_c^*$ (we have data 
for several values of $z_c$ with step $\Delta z_c = 0.1$). 
If we apply instead the 
equal-height method we obtain $z_c^*(c) R_c^3 \approx 129.3$ and 
$z_c^*(p) R_c^3 \approx 127.5$ from the two distributions.
The colloid packing fractions at coexistence are therefore
\begin{equation}
\begin{array}{lll}
z_c^*(c) R_c^3 = 129.3 &\qquad \eta_{c,\rm gas} \approx 0.039 & \qquad
                    \eta_{c,\rm liq} \approx 0.294,
\\
z_c^*(p) R_c^3 = 127.5 &\qquad \eta_{c,\rm gas} \approx 0.038 & \qquad
                    \eta_{c,\rm liq} \approx 0.292.
\end{array}
\end{equation}
The results are very close with each other and consistent with those 
reported in Eq.~(\ref{coexdata-1}). 
Similar conclusions are obtained for the 
polymer packing fractions at coexistence.
For $f = 0.7$, $R_{\rm dis}/R_c = 1.0$, $\eta_r^p \approx 1.82$
($z_p R_c^3= 0.85$), and $L/R_c = 16$, the case reported in 
Fig.~\ref{critical-fugacity2}, we obtain
$z_c^*(c) R_c^3 \approx 454$ and
$z_c^*(p) R_c^3 \approx 440$ from the two distributions
(equal-height method). 
At coexistence we find then
\begin{equation}
\begin{array}{lll}
z_c^*(c) R_c^3 = 454 &\qquad \eta_{c,\rm gas} \approx 0.096 & \qquad
                    \eta_{c,\rm liq} \approx 0.257,
\\
z_c^*(p) R_c^3 = 440 &\qquad \eta_{c,\rm gas} \approx 0.100 & \qquad
                    \eta_{c,\rm liq} \approx 0.258.
\end{array}
\end{equation}
Even though the estimates of the coexistence colloid fugacity differ somewhat,
the two estimates of the colloid packing fractions at coexistence 
are quite close.

\subsection{Sample-to-sample fluctuations} \label{sec3.3}

\begin{figure}
\begin{tabular}{ccc}
\epsfig{file=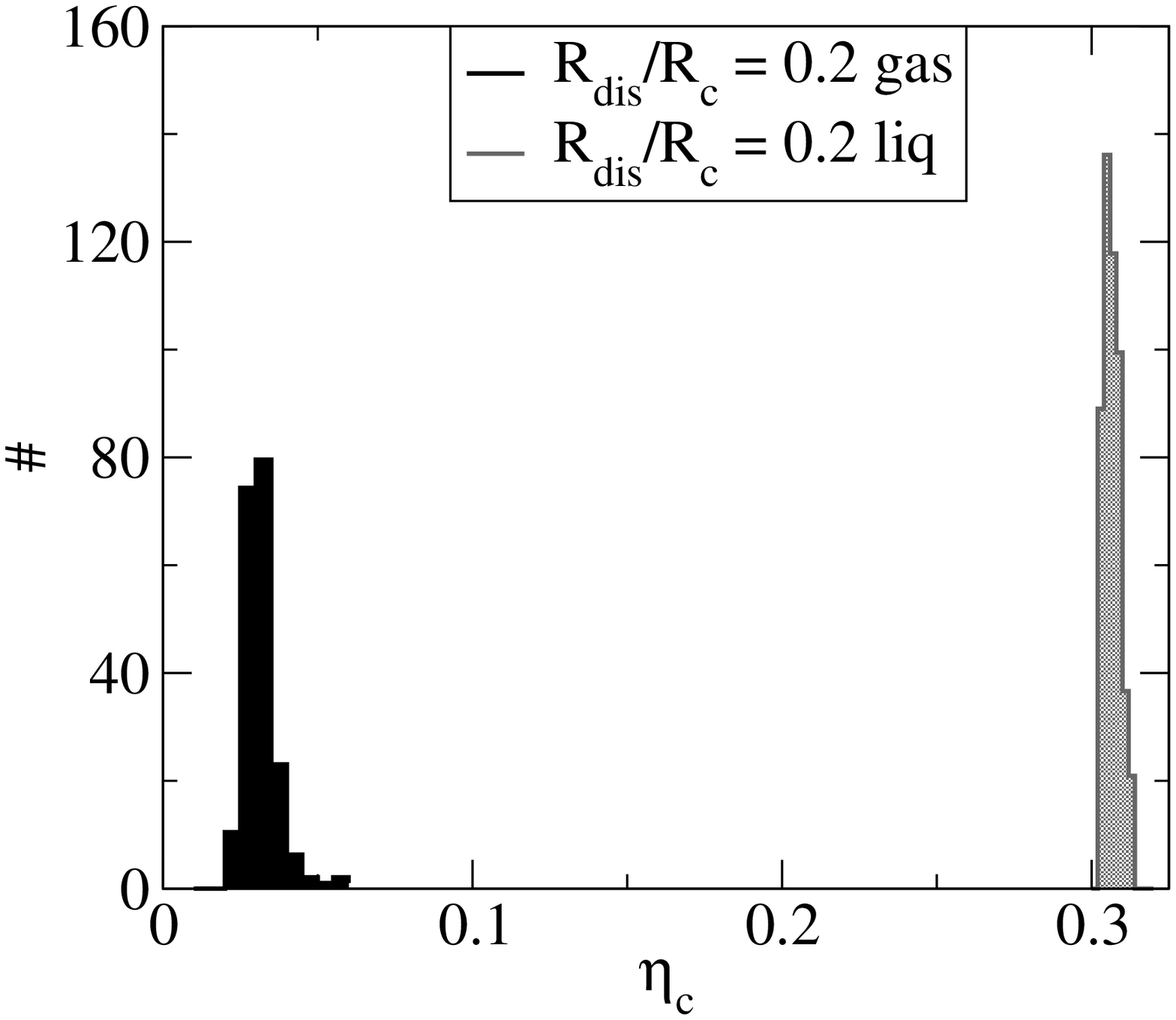,angle=0,width=5.5truecm} &
\epsfig{file=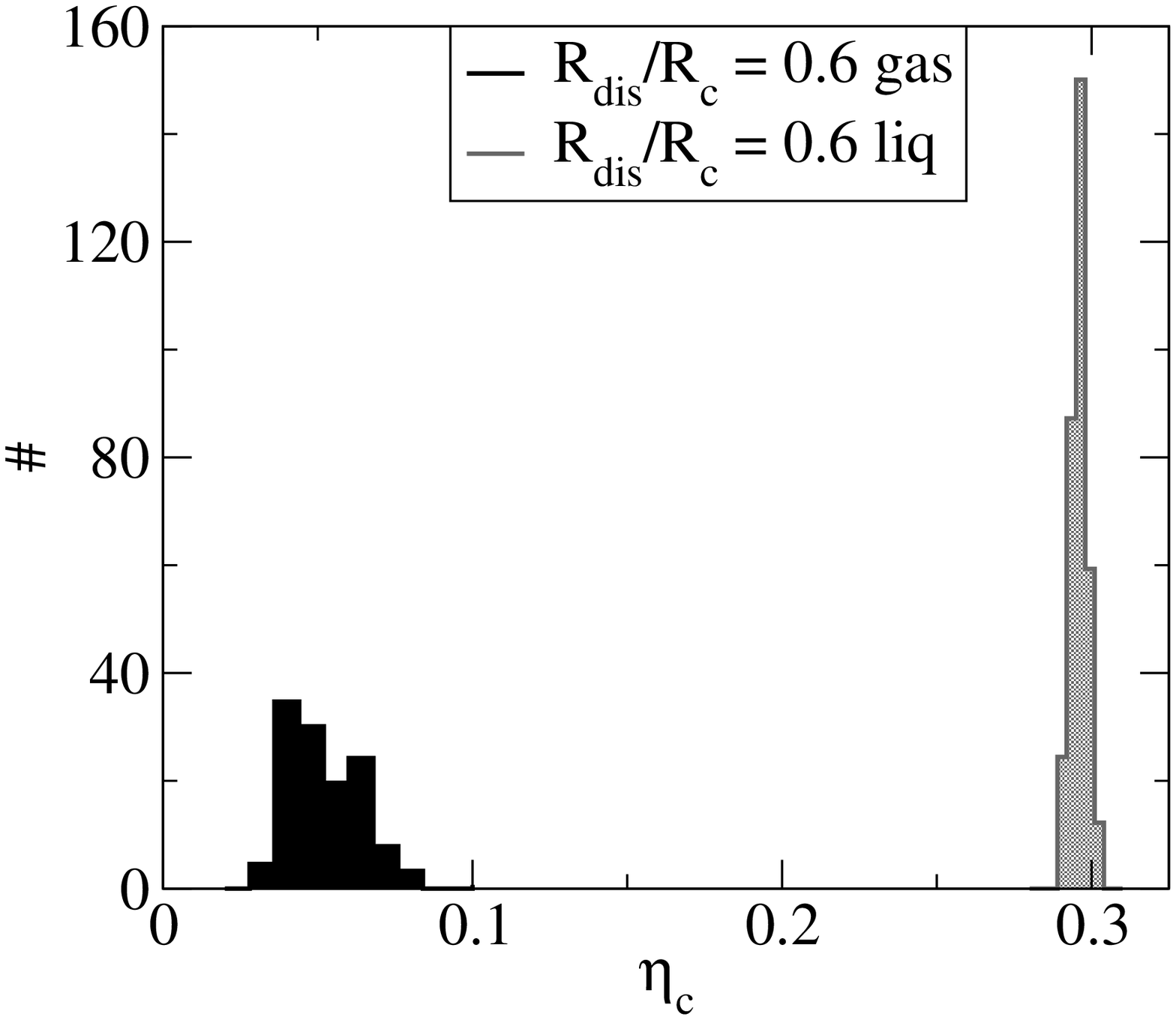,angle=0,width=5.5truecm} &
\epsfig{file=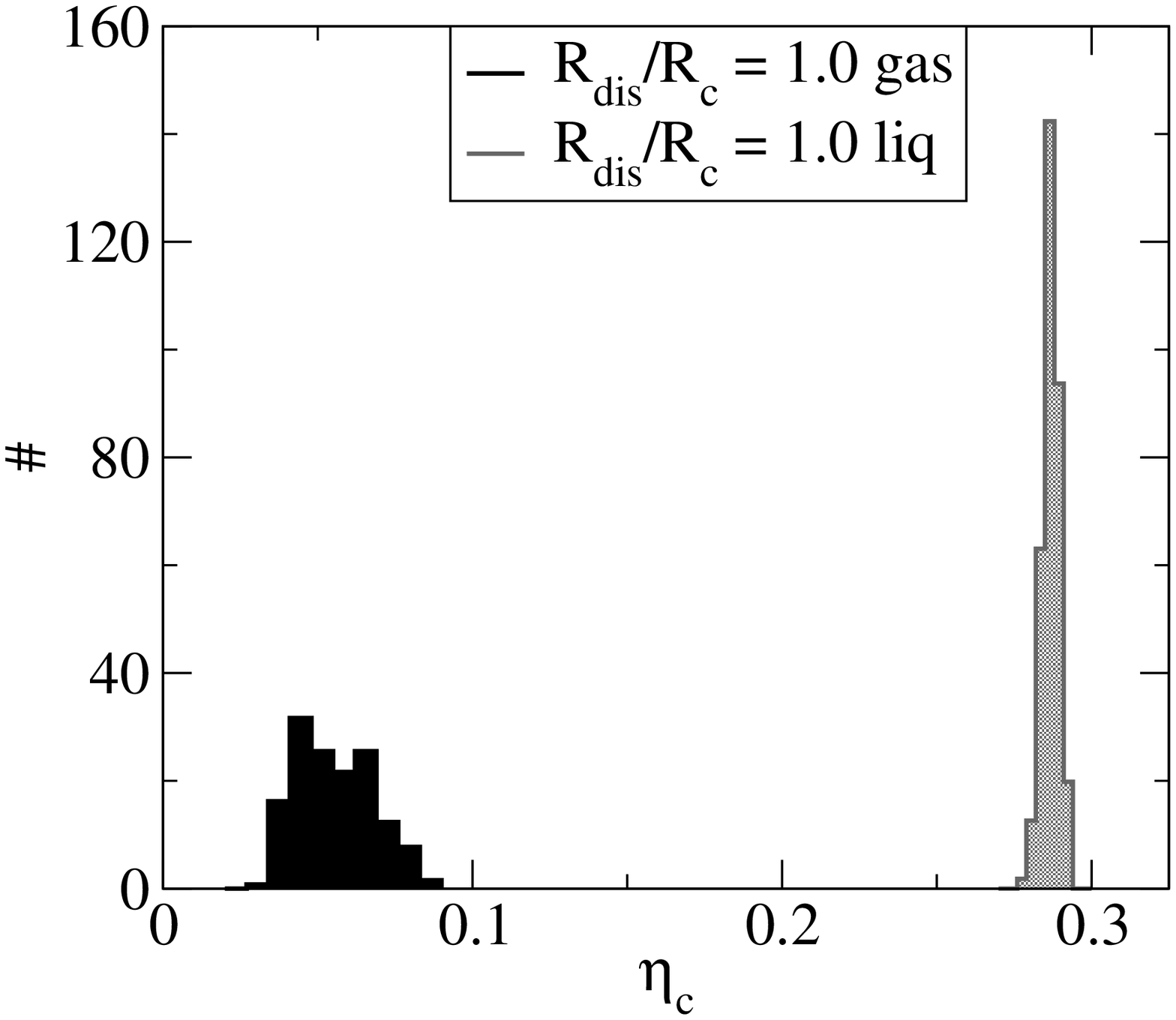,angle=0,width=5.5truecm} \\
\epsfig{file=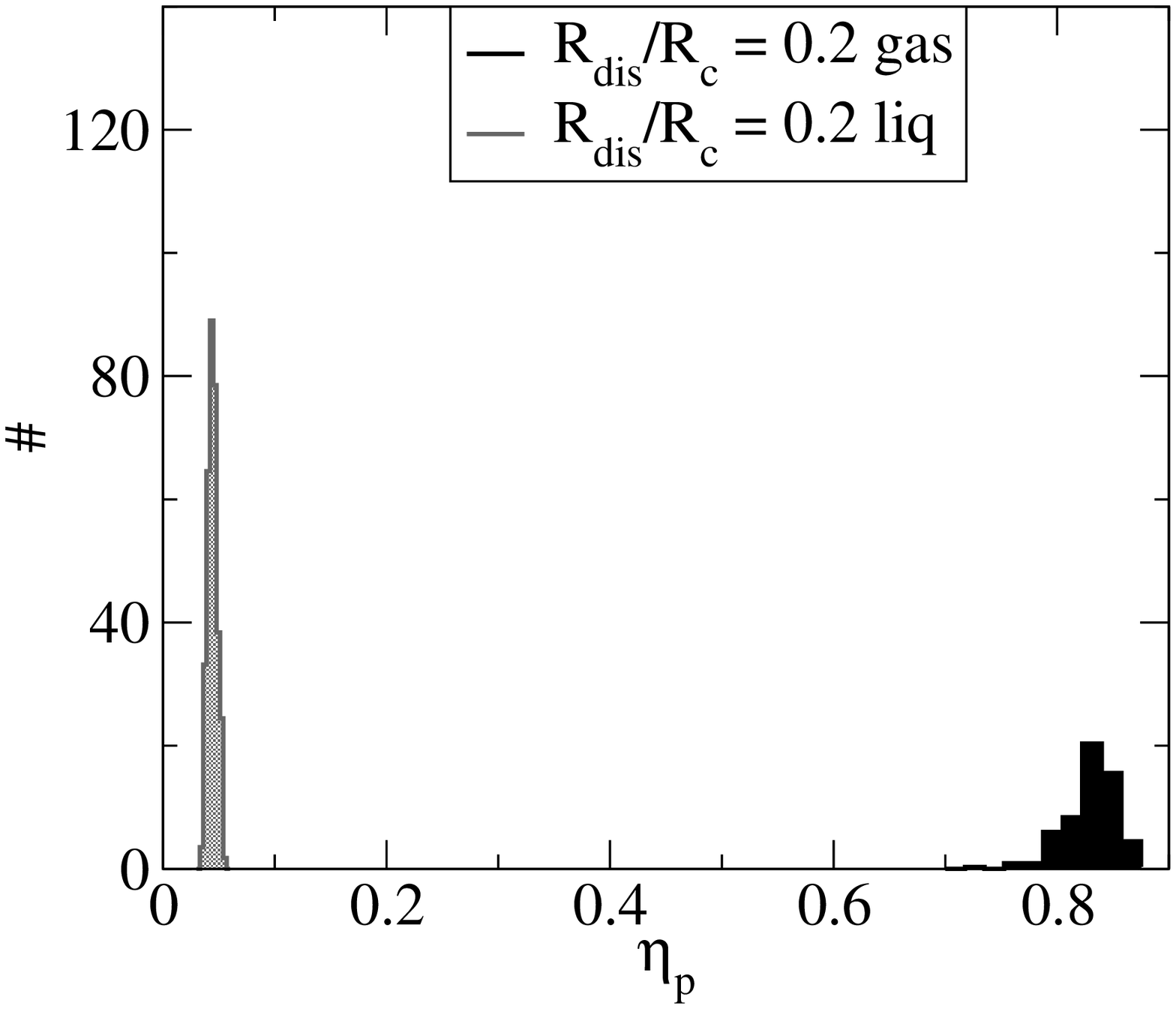,angle=0,width=5.5truecm} &
\epsfig{file=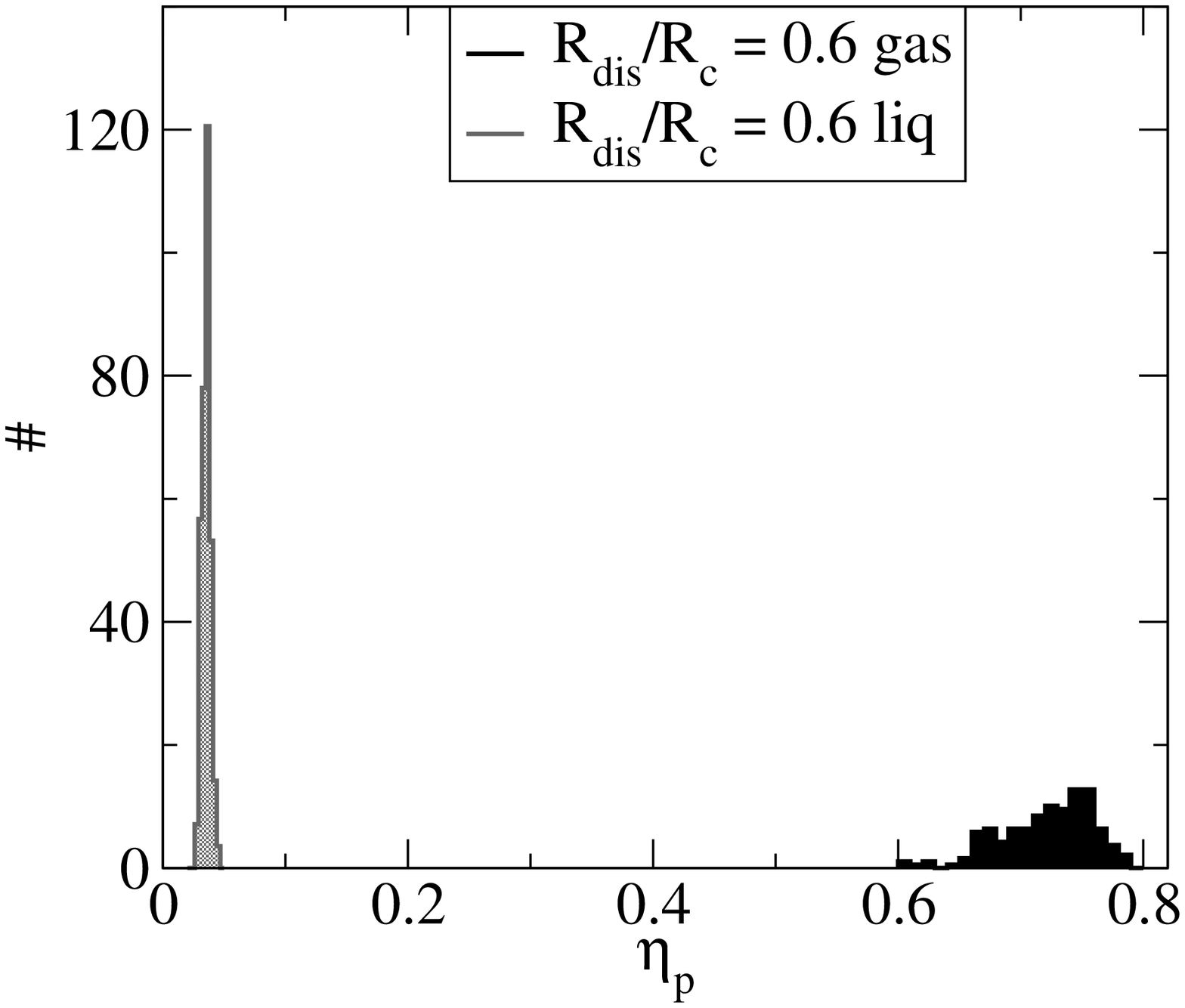,angle=0,width=5.5truecm} &
\epsfig{file=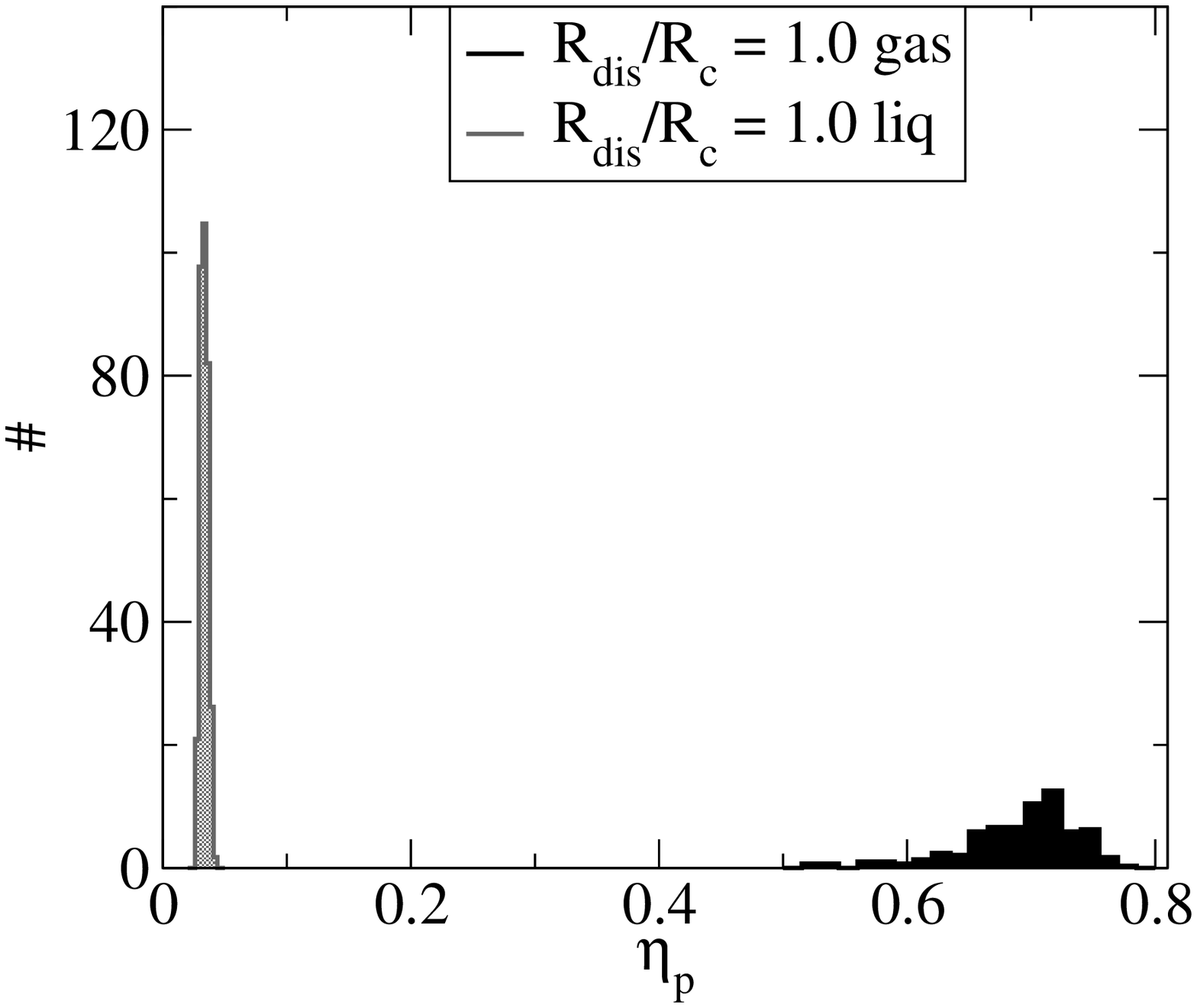,angle=0,width=5.5truecm} \\
\end{tabular}
\vspace{1cm}
\caption{(Color online)  Distribution of  
the coexisting colloid (top) and polymer (bottom) 
packing fractions for $\eta_p^r \approx 1.24$
($z_p R_c^3 = 0.58$), $f = 0.4$ and three different values of 
$R_{\rm dis}/R_c$: 0.2, 0.6, 1.0. 
The colloid-gas phase data are reported in 
black, while the colloid-liquid phase data are reported in gray.
Data from simulations with $L/R_c=16$.
}
\label{fig-distributionepicchi}
\end{figure}    

\begin{figure}
\begin{tabular}{cc}
\epsfig{file=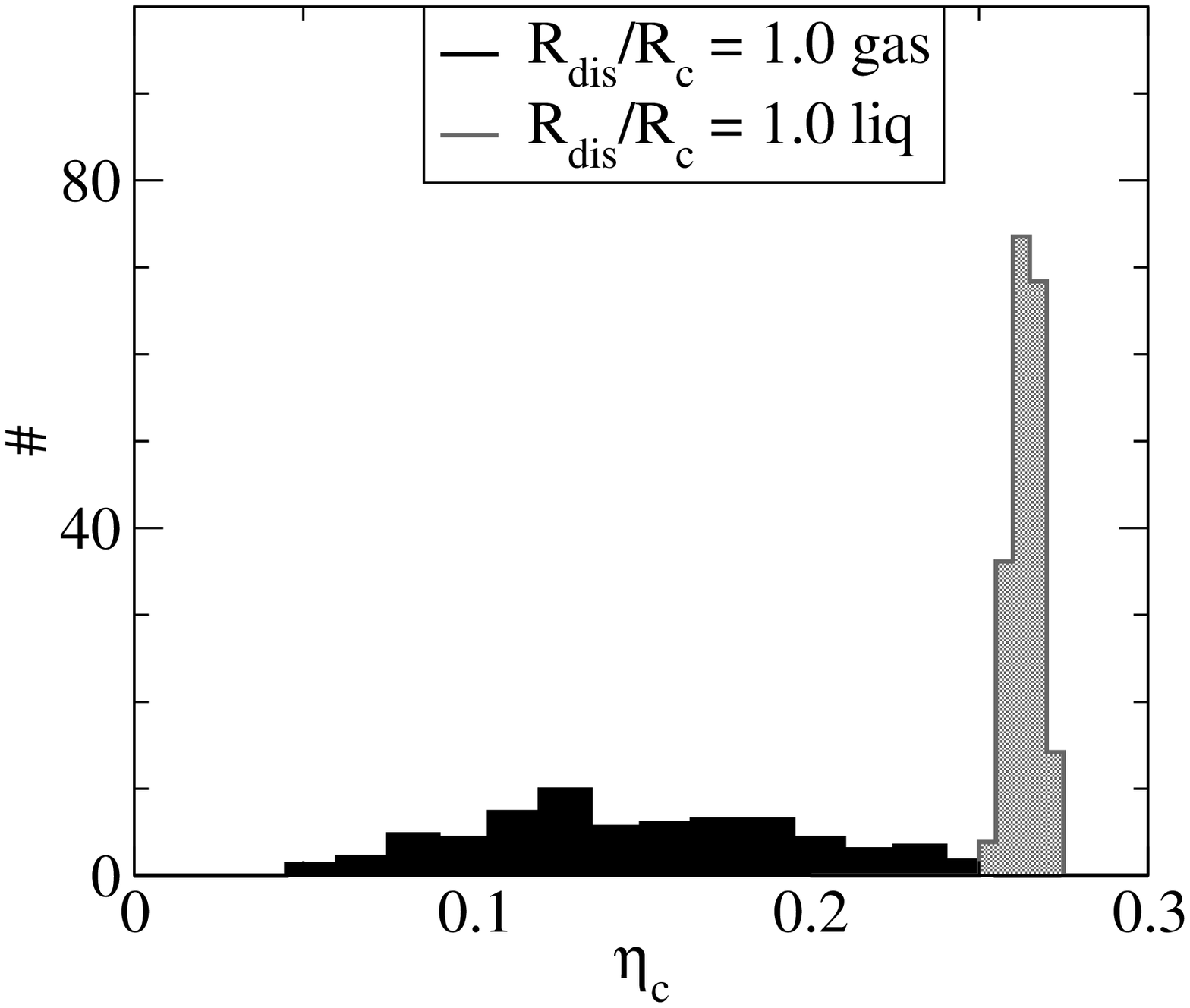,angle=0,width=8truecm} &
\epsfig{file=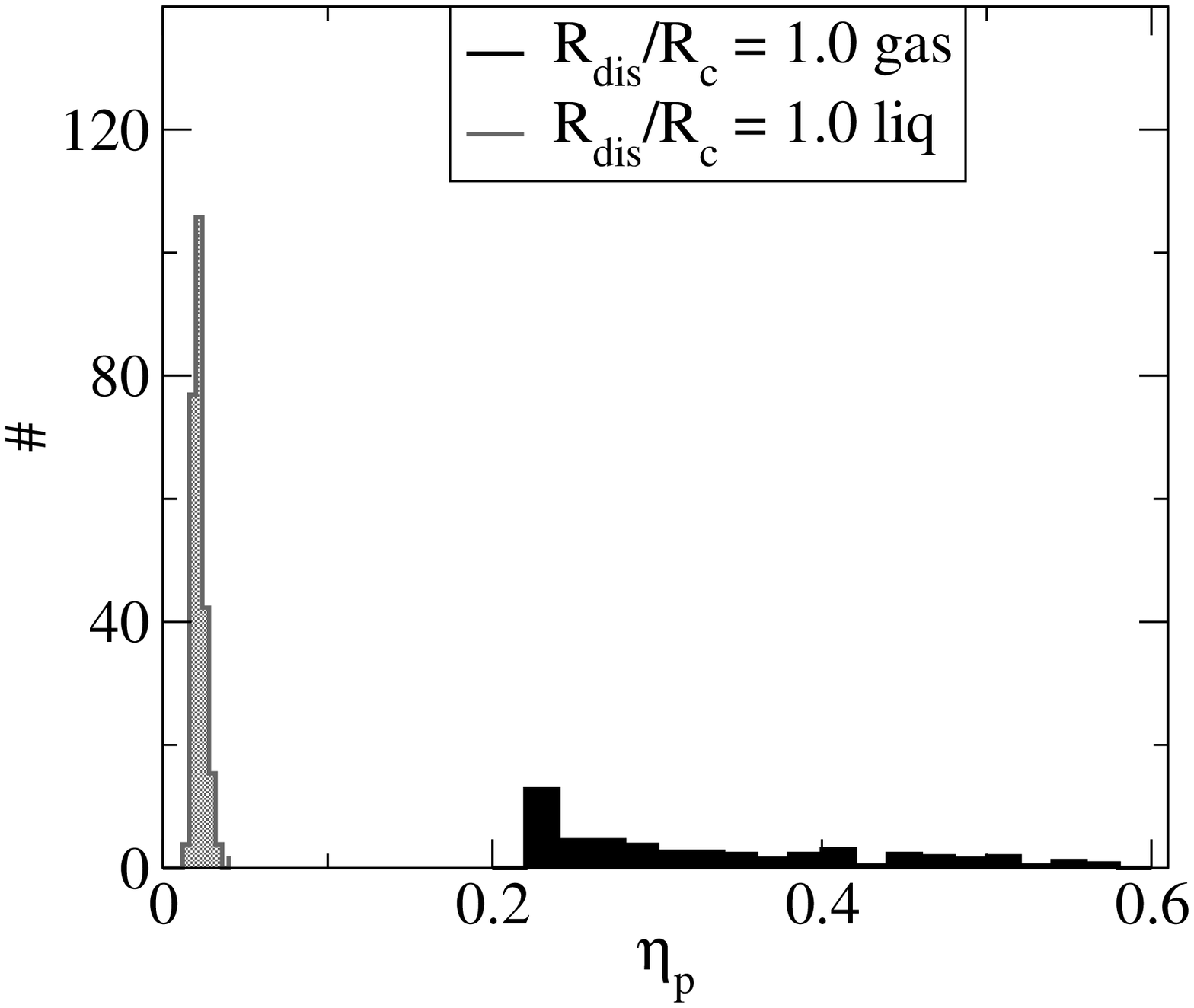,angle=0,width=8truecm} \\
\end{tabular}
\vspace{1cm}
\caption{Distribution of  
the coexisting colloid (left) and polymer (right) 
packing fraction for $\eta_p^r \approx 1.82$
($z_p R_c^3 = 0.85$), $f = 0.7$ and 
$R_{\rm dis}/R_c = 1$.
The colloid-gas phase data are reported in 
black, while the colloid-liquid phase data are reported in gray. 
Data from simulations with $L/R_c=16$.
}
\label{fig-distributionepicchi2}
\end{figure}    

It is interesting to understand how the results obtained from the 
sample average compare with those that would be obtained by 
determining the coexisting phases for each disorder realization. 
In Fig~\ref{fig-distributionepicchi} we report the distributions of the 
colloid and polymer packing fractions at coexistence computed from
each disorder configuration. The distributions of the packing fractions
corresponding to the colloid-liquid phase are very narrow and are
centered at the value obtained from the analysis of the average distributions.
On the other hand, the distributions for the colloid-gas phase are broad,
especially for $R_{\rm dis}/R_c = 0.6$ and 1. Since the broadness of the 
distribution is a finite-size effect --- we expect the width of the 
distributions to scale as $1/\sqrt{L}$ as $L\to\infty$ --- this is an
indication that we should expect some finite-size dependence on 
our determination of the colloid-gas branch of the coexistence curves. 
The results reported in the next section confirm these 
expectations. 

Finally, we also report the distributions for the case $f = 0.7$, 
$R_{\rm dis}/R_c = 1$, $\eta_p^r \approx 1.82$ ($z_p R_c^3 = 0.85$)
(the average colloid-number histograms
are reported in Fig.~\ref{critical-fugacity2}), for which we have been
unable to determine the coexistence fugacity using the equal-area method. 
The distribution of the colloid and polymer packing fractions at
coexistence are reported in Fig.~\ref{fig-distributionepicchi2}
and clearly explain the origin of the difficulties. The position of the 
colloid-gas branch varies significantly from sample to sample.
We are thus far from the infinite-volume limit, since in this limit
sample fluctuations are expected to disappear except close to the 
critical point. These results provide again evidence that 
size effects are large for these values of the parameters.

\subsection{Finite-size effects} \label{sec3.4}

\begin{figure}
\begin{tabular}{ll}
\epsfig{file=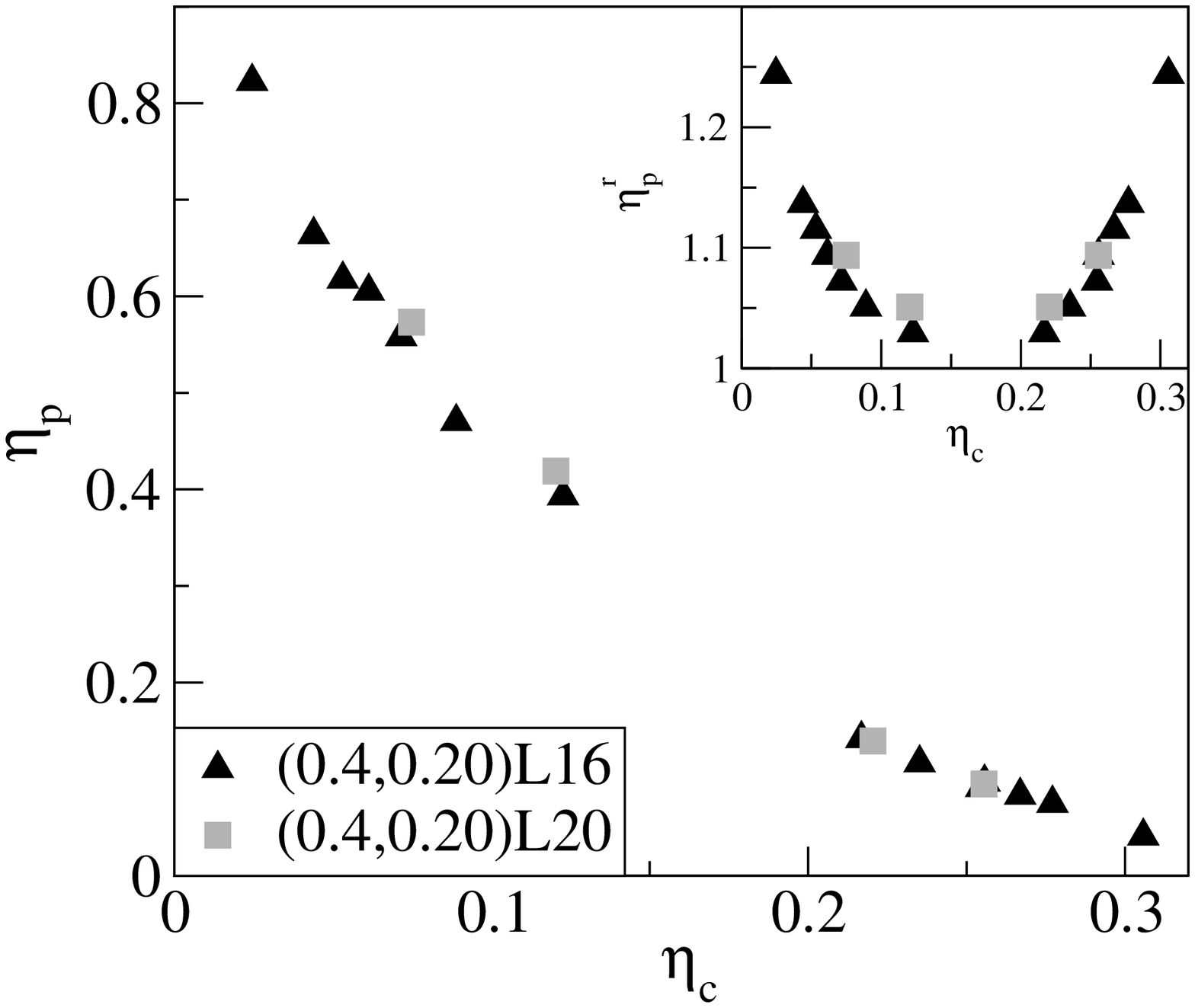,angle=0,width=8truecm} 
  \hspace{-0.0truecm} &
\epsfig{file=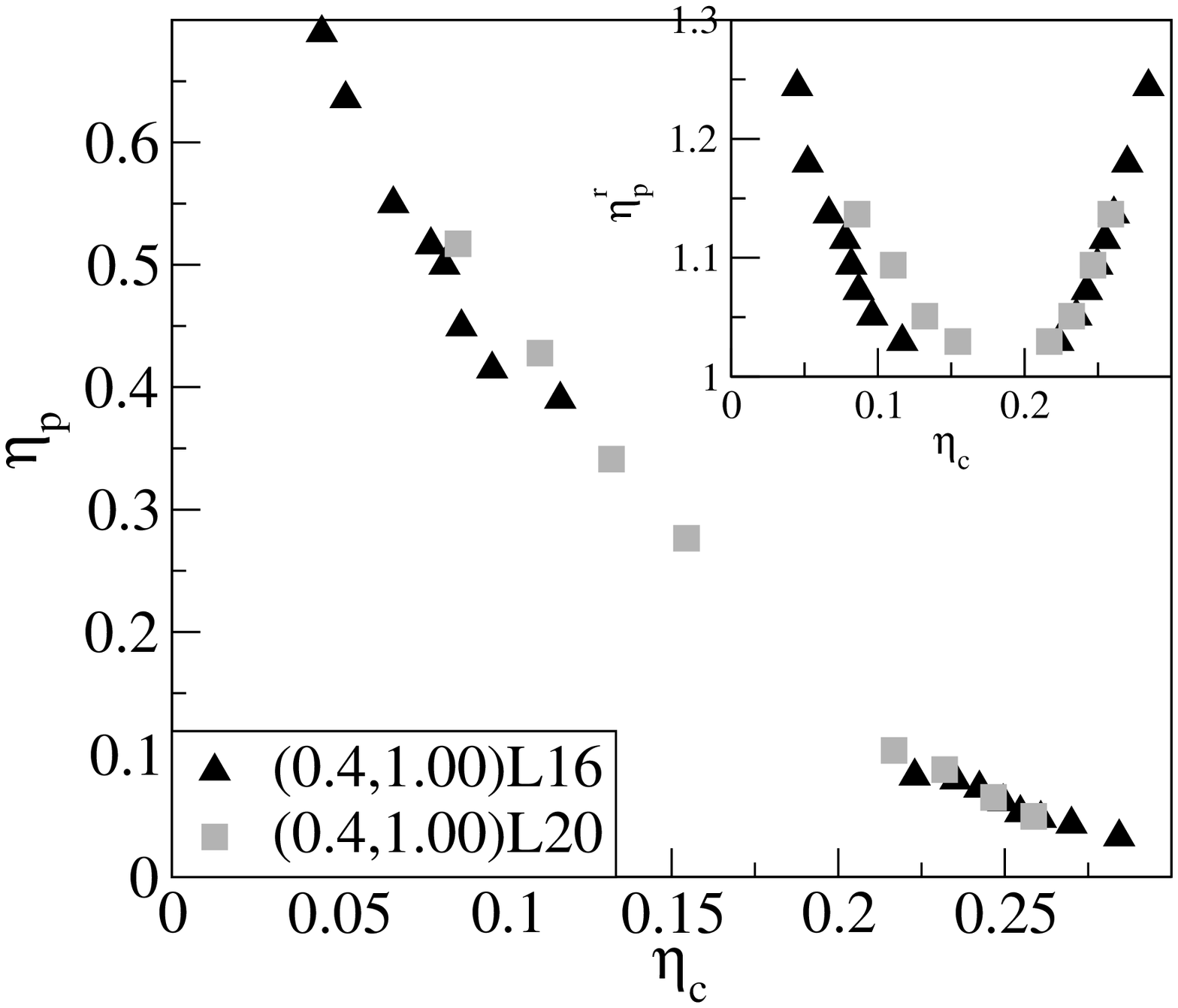,angle=0,width=8truecm} \\
\vspace{0.2cm} \\
\epsfig{file=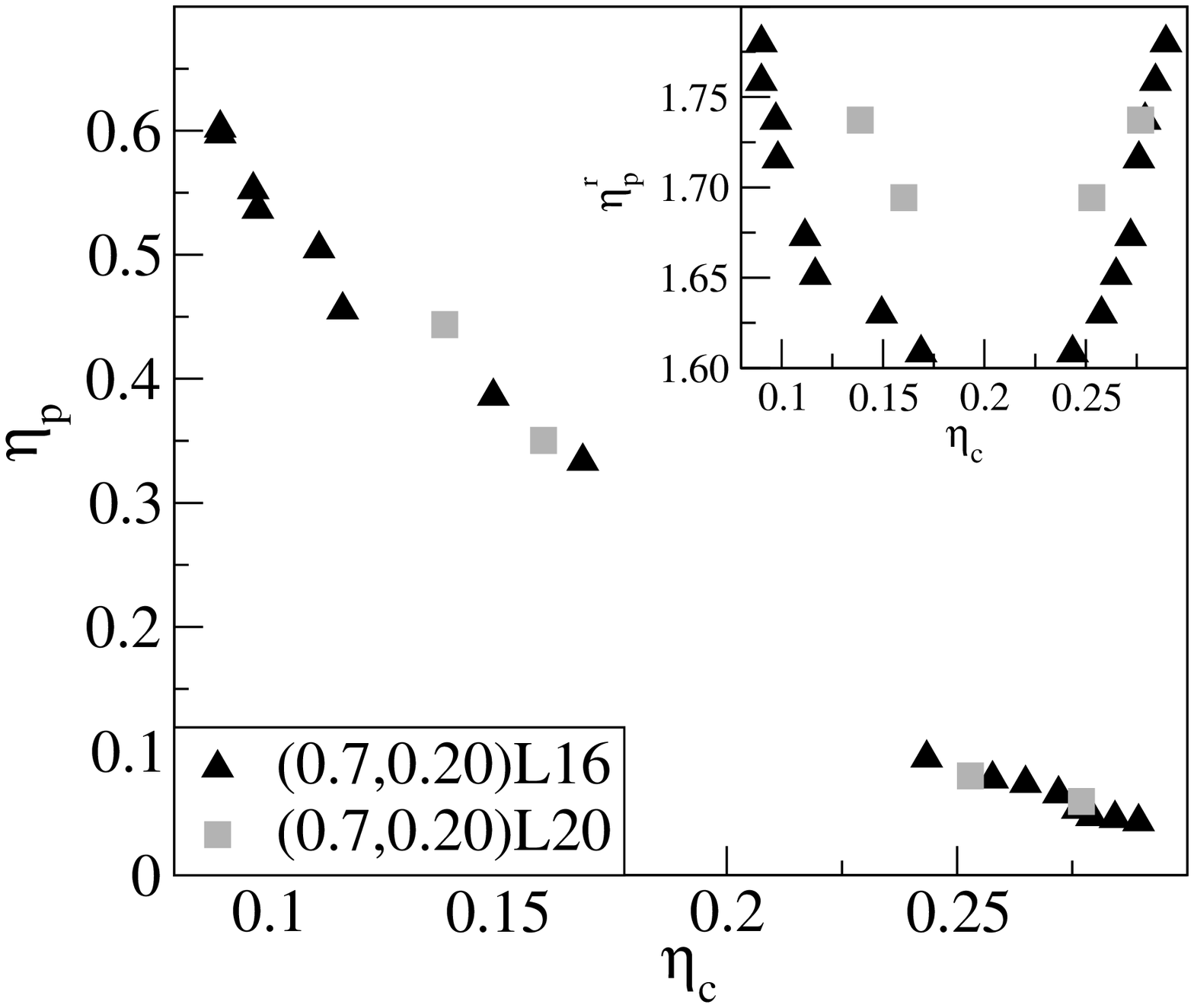,angle=0,width=8truecm} 
  \hspace{-0.0truecm} &
\epsfig{file=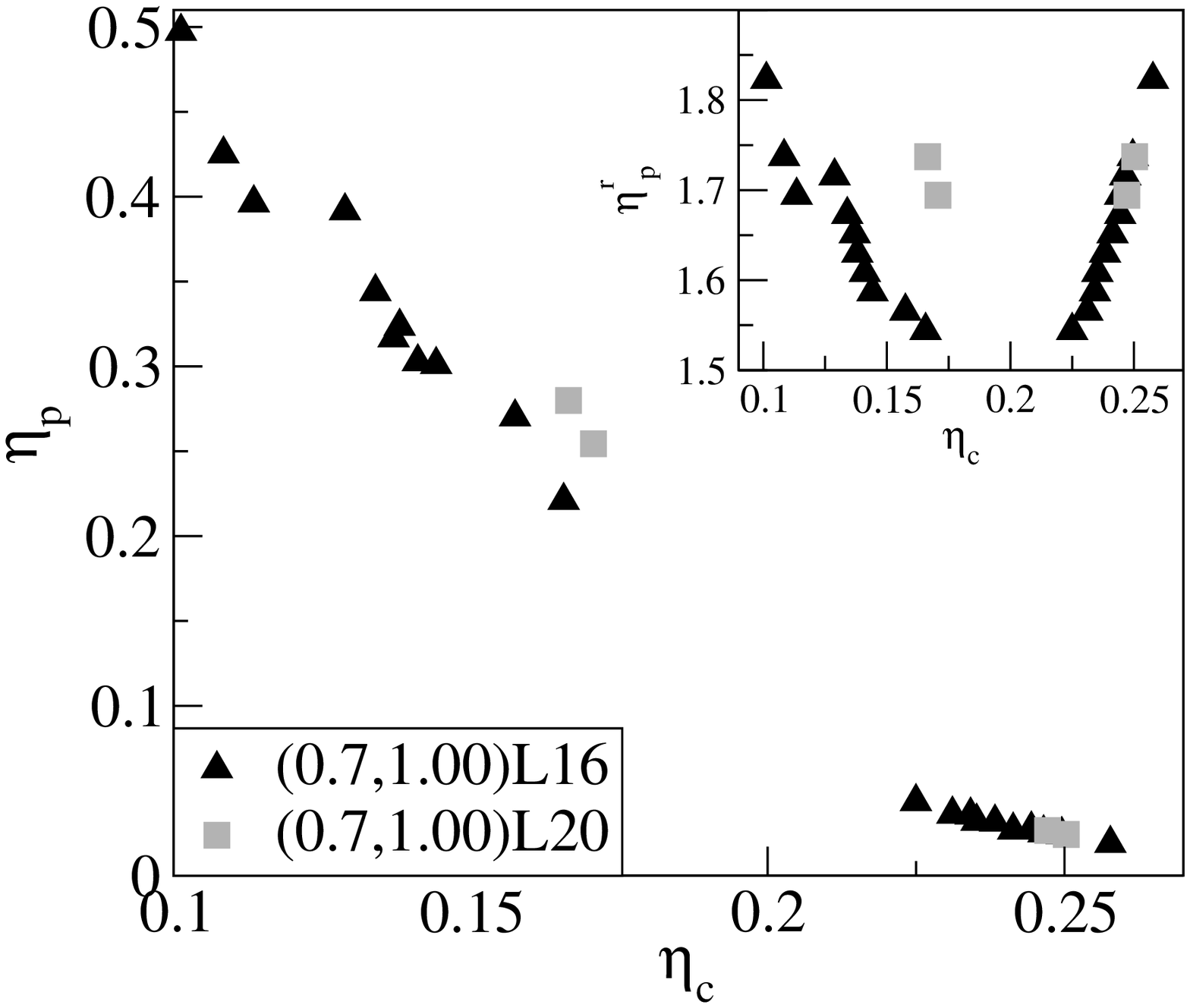,angle=0,width=8truecm} \\
\vspace{0.0cm}
\end{tabular}
\caption{Fluid-fluid binodal curves for 
  $f = 0.4$, $R_{\rm dis}/R_c = 0.2$ (top left), 
  $f = 0.4$, $R_{\rm dis}/R_c = 1.0$ (top right), 
  $f = 0.7$, $R_{\rm dis}/R_c = 0.2$ (bottom left), 
  $f = 0.7$, $R_{\rm dis}/R_c = 1.0$ (bottom right).
  We report the results for $L=16R_c$ and $L=20R_c$ 
  in the $\eta_c$, $\eta_p$ plane
  and (inset) in the reservoir representation $\eta_c$, $\eta_p^r$.
}
\label{binodali-L}
\end{figure}    

The analyses presented above show that size effects may still be relevant 
for the data with $L/R_c = 16$. They appear to increase with
increasing $f$ and/or $R_{\rm dis}/R_c$ and should be particularly
large for $f = 0.7$ and $R_{\rm dis}/R_c = 1.0$. To investigate size effects
we have performed additional simulations with $L/R_c = 20$.
In Fig.~\ref{binodali-L} we compare the results for the coexistence curve
obtained by using these two different box sizes. 
We report results both in terms of 
$\eta_c$ and $\eta_p$ and also in the reservoir representation (see insets)
in terms of $\eta_c$ and $\eta_p^r$.

On top we show the results for $f = 0.4$. Corrections here appear to be 
under control: the plots in the terms of $\eta_c$ and $\eta_p$ show a small
size dependence, while those in the reservoir representation appear to 
be reliable except close to the critical point, which is not unexpected
since size corrections are large in a neighborhood of a second-order 
phase transition. For $f = 0.7$, the colloid-liquid branch 
is determined quite reliably. On the other hand, the colloid-gas phase boundary
varies significantly when $L$ changes, especially for the case 
$R_{\rm dis}/R_c = 1.0$. This is not unexpected, given the results shown
in the previous sections. Indeed, in all cases the polymer and colloid 
histograms are characterized by very narrow colloid-liquid peaks whose
positions have a tiny dependence on the fugacity 
$z_c$, so that, even if $z_c^*$ is not precisely determined,
the determined values $\eta_{c,\rm liq}$ and $\eta_{p,\rm liq}$ are 
quite reliable. In the colloid-gas phase the distributions are instead
very broad, a clear indication that we are far from the infinite-volume limit.
As we have already remarked, it is also possible that, for some values of the 
parameters, the system is in the one-phase region, even if the finite-size
data show double peaks.

\subsection{Demixing curves in the reservoir representation} \label{sec3.5}

\begin{figure}
\begin{tabular}{ll}
\epsfig{file=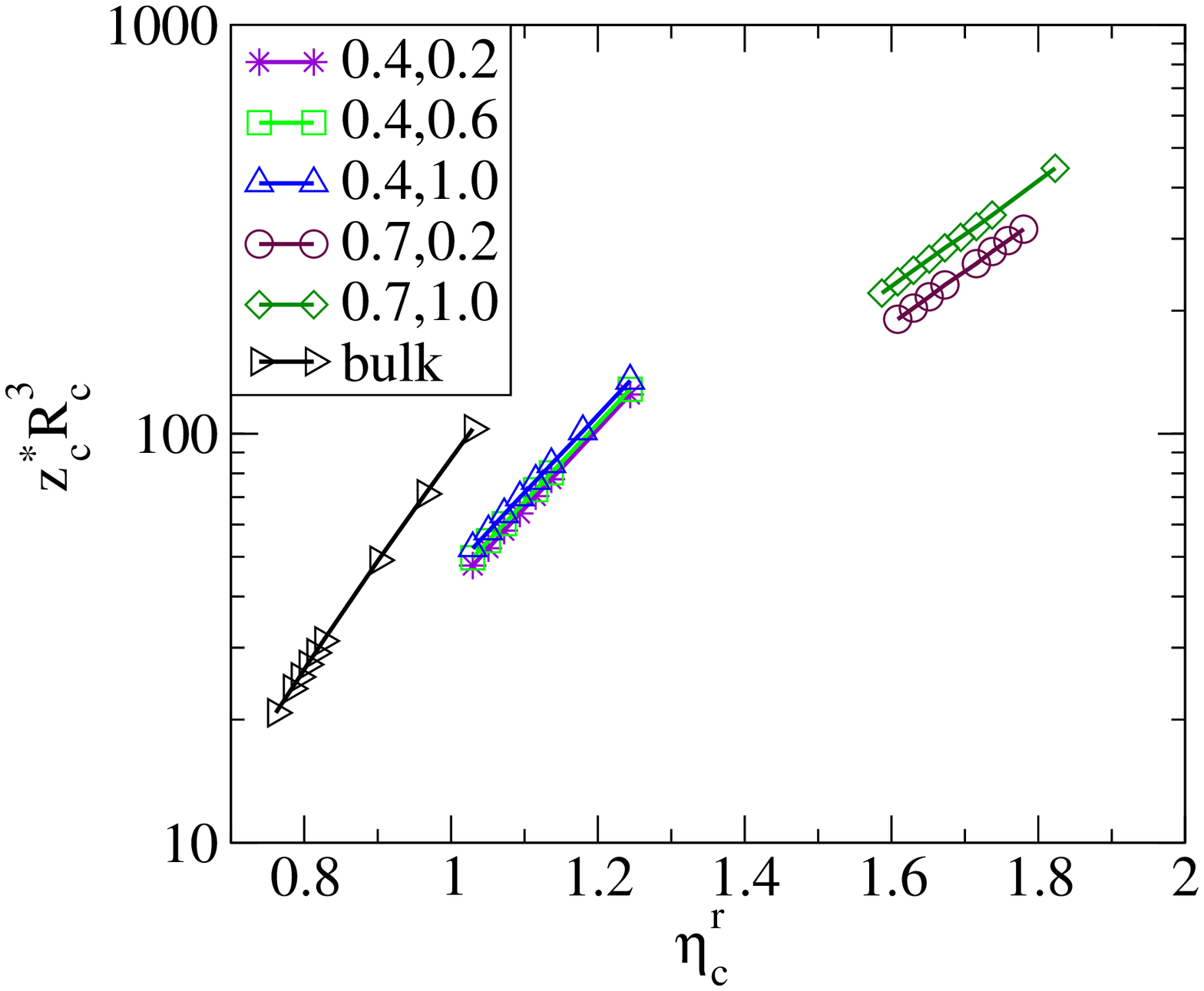,angle=0,width=7truecm} & 
\epsfig{file=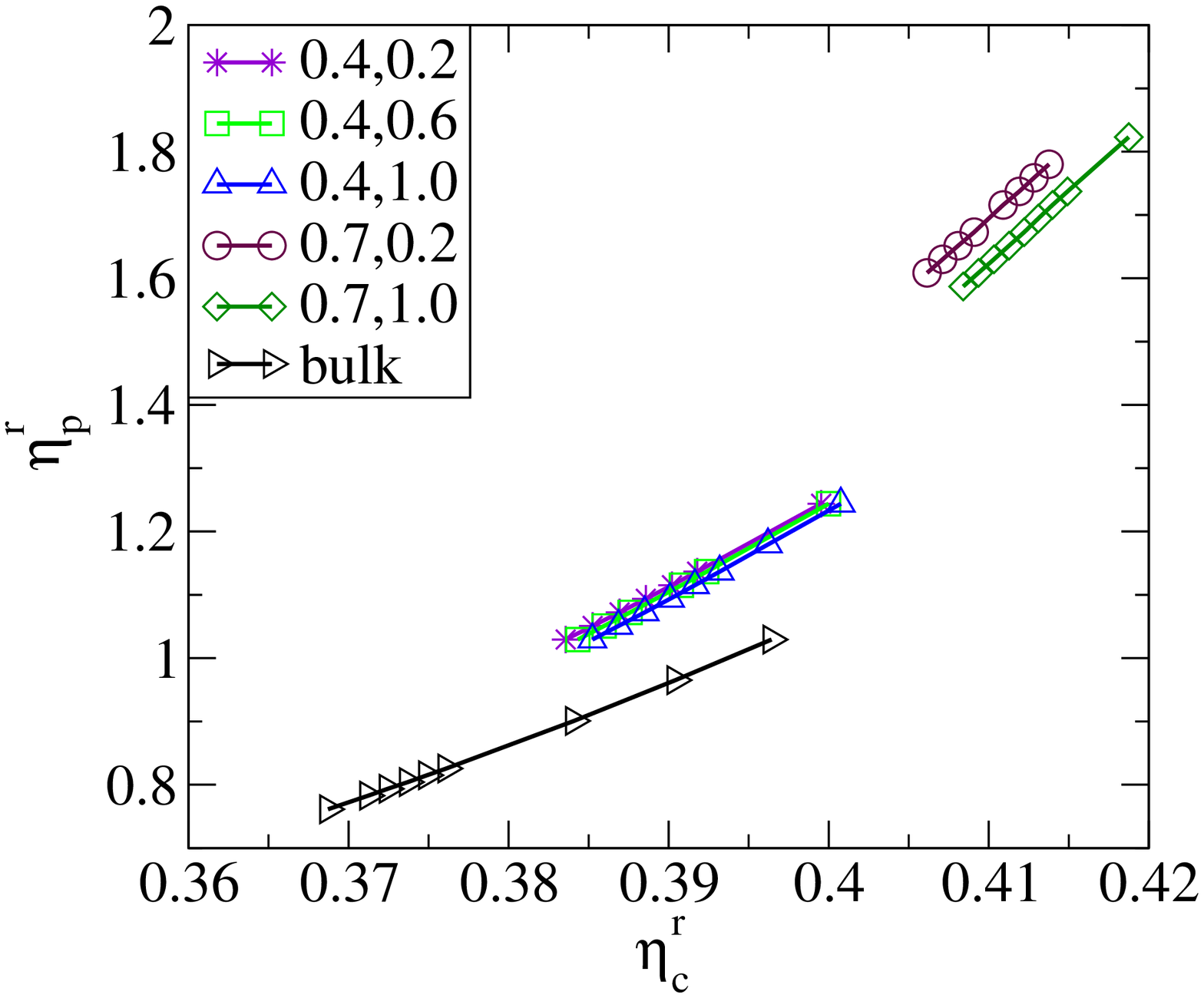,angle=0,width=7truecm}  \\
\end{tabular}
\vspace{0cm}
\caption{Estimates of $z^*_c R_c^3$ 
as a function of the 
polymer reservoir packing fraction $\eta_p^r$ (left) and of 
$\eta_p^{r*}$ at coexistence in terms of the 
colloid reservoir packing fraction $\eta_c^r$ (right).  
Results for $L/R_c = 16$.
In the legend the first number corresponds to $f$, while the second one 
gives the ratio $R_{\rm dis}/R_c$.
}
\label{fig-zstar}
\end{figure}

The estimates of $z_c^*$ [we report the average of $z_c^*(c)$ and
$z_c^*(p)$] as a function of $\eta^r_p$ for $L/R_c = 16$ 
are reported in Fig.~\ref{fig-zstar}.  
In order to compare our results 
with those of Ref.~\cite{SSKK-02}, we also report the estimates 
of the polymer reservoir packing fraction $\eta_p^{r*}$ at coexistence in
terms of the colloid reservoir packing fraction $\eta_c^{r}$
\cite{footnote-etacr}.
Note that, on a logarithmic 
scale, the values $z_c^* R_c^3$ for each $f$ and $R_{\rm dis}/R_c$ 
lie quite precisely on a straight line, indicating that the 
colloid chemical potential at coexistence is well approximated by a linear 
function in $\eta_p^r$. Moreover, the position of the demixing curve 
depends essentially only on $f$. The ratio $R_{\rm dis}/R_c$, hence 
the topological structure of the matrix, does not change significantly 
the coexistence curve. We have not performed a careful 
finite-size scaling analysis close to the critical point 
(a detailed discussion of the methods appropriate for 
random-field Ising critical points is reported in 
Refs.~\cite{VBL-08,VTB-10,FV-11b}) and thus 
we are not able to estimate $\eta_{p,\rm crit}^r$ and 
$\eta^r_{c,\rm crit}$ precisely. We only note that for $L/R_c = 16$ 
and $L/R_c = 20$
double peaks are observed only for $\eta_p^r \gtrsim 1.00$, 1.03
for all three values of $R_{\rm dis}/R_c$.
We can thus set the lower bounds
$\eta_{p,\rm crit}^r \gtrsim 1.03$ and $\eta^r_{c,\rm crit} \gtrsim 0.38$.
For $f = 0.7$ size effects are significantly larger than for $f = 0.4$,
but we can still obtain the bounds 
$\eta_{p,\rm crit}^r \gtrsim 1.6$, $\eta^r_{c,\rm crit} \gtrsim 0.405$. 
 
We can use our results to understand qualitatively the behavior 
of a bulk colloid-polymer mixture in chemical equilibrium 
with the same dispersion adsorbed in a porous matrix. 
The main question is whether one can observe different phases 
in the bulk and in the matrix. 
If we use $\eta_c^r$ as control parameter, we see that for 
$\eta_c^r < \eta_{c,\rm crit,bulk}^r\approx 0.37$ 
there is no transition, neither 
in the bulk nor in the matrix. If $\eta_c^r$ is larger, one may have 
a transition in the bulk and no transition in the matrix, given that 
$\eta_{c,\rm crit}^r$ increases with $f$. For instance, for $f = 0.4$ 
and $0.37\lesssim \eta_c^r \lesssim 0.385$
we only observe phase separation in the bulk. 
If we increase further  $\eta_c^r$ ($\eta_c^r \gtrsim 0.385$ for $f = 0.4$) 
the behavior is more complex, since phase separation occurs both 
in the bulk and in the matrix. For small $\eta_p^r$  
the mixture is in the colloid-liquid phase both in the bulk and in the matrix. 
If $\eta_p^r$ is increased, that is polymers are added,
the bulk coexistence curve is reached, see 
Fig.~\ref{fig-zstar}. Above the demixing line,
one observes two different phases: in the bulk the system is in the 
colloid-gas phase, while in the matrix a colloid-liquid phase occurs.
Thus, the presence of the matrix may induce, for certain values of the 
parameters, a capillary condensation of the colloids.  
Finally, for large $\eta_p^r$ above the matrix coexistence curve, 
a colloid-gas phase occurs both in the 
bulk and in the matrix.

\subsection{Binodals in the system representation} \label{sec3.6}

\begin{figure}
\begin{tabular}{cc}
\epsfig{file=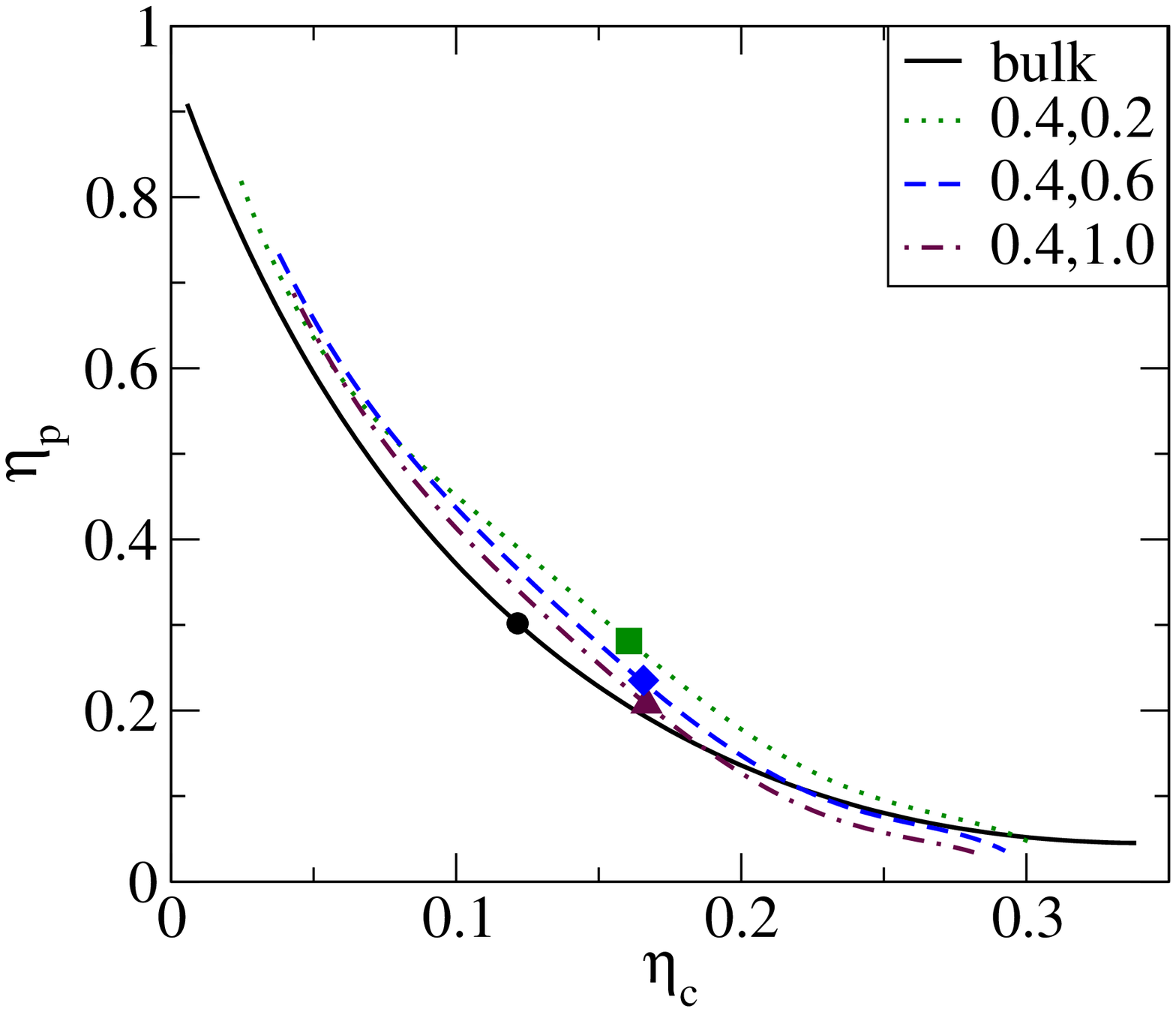,angle=0,width=8truecm} &
\epsfig{file=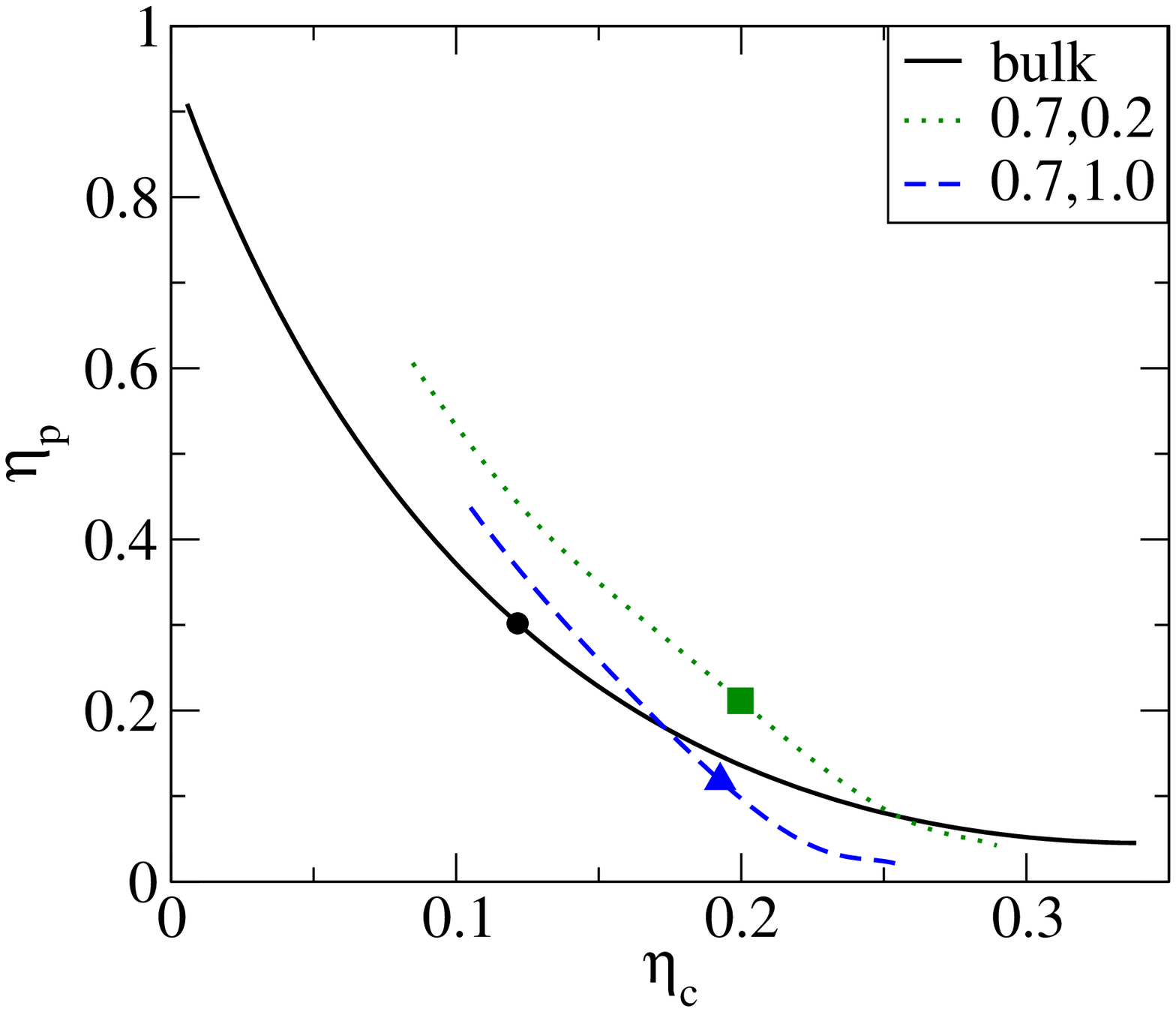,angle=0,width=8truecm} 
 \\
\end{tabular}
\vspace{0cm}
\caption{Binodal curves (obtained by interpolating the 
  data reported in Fig.~\ref{binodali-L}) for $f = 0.40$ (left) and 
  $f = 0.70$ (right). In both plots we also report the bulk 
  binodal curve and the effective critical points.
In the legend the first number corresponds to $f$, while the second one 
gives the ratio $R_{\rm dis}/R_c$.
  Simulations in a box of size $L=16R_c$.
}
\label{binodali-finali}
\end{figure}    

In Fig.~\ref{binodali-finali}
we report the results for the binodals (we interpolate the 
MC data for $L/R_c = 16$) as a function of 
$\eta_c$ and $\eta_p$. For $f = 0.4$ they are quite close to the bulk binodal
curve and show only a tiny dependence on the 
ratio $R_{\rm dis}/R_c$. In the figure we also report an estimate of the 
critical point obtained by determining the intersection of the diameter
with the interpolation of the coexistence data. This provides a very rough 
estimate of the critical parameters, which can only be accurately determined 
by performing a careful finite-size scaling analysis. In the bulk the 
analysis of the results with $L/R_c = 16$ gives
$\eta_{c,\rm crit} \approx 0.12$, 
$\eta_{p,\rm crit} \approx 0.31$, which should be compared  
with the precise determination \cite{VH-04bis,VHB-05}
\begin{equation}
\eta_{c,\rm crit} = 0.1340(2), \qquad\qquad \eta_{p,\rm crit} = 0.3562(6).
\end{equation}
Apparently our simple extrapolation underestimates $\eta_{c,\rm crit}$ by 10\%
and $\eta_{p,\rm crit}$ by 15\%, which we can take as indications of the 
systematic error.
If we perform the same analysis for $f = 0.4$ we obtain 
$\eta_{c,\rm crit} \approx 0.17$ 
for all values of $R_{\rm dis}/R_c$:
the colloid packing fraction at criticality is essentially independent
of the matrix topology and increases with increasing $f$. 
As for the polymer packing fraction, the dependence on the size ratio
$R_{\rm dis}/R_c$ is somewhat larger. The value $\eta_{p,\rm crit}$ decreases
with increasing $R_{\rm dis}/R_c$ and varies between 0.23 and 0.29. 
If we assume that the results for $L/R_c = 16$ underestimate the 
correct, infinite-volume results by 10\% and 15\% as they do in the bulk, 
we would guess $\eta_{c,\rm crit} \approx 0.19$ and 
$0.27 \lesssim \eta_{p,\rm crit} \lesssim 0.34$. 

As we have already stated our results for $f = 0.7$ are only reliable 
in the colloid-liquid phase. In this regime the binodal curve depends somewhat
on $R_{\rm dis}/R_c$: for $R_{\rm dis}/R_c = 0.2$ it is 
close to the bulk curve, while for $R_{\rm dis}/R_c = 1$ it is significantly
below it. The critical point position is consistent with what was 
observed for $f = 0.4$: $\eta_{c,\rm crit}$ shows little dependence on 
$R_{\rm dis}/R_c$, while $\eta_{p,\rm crit}$ decreases with 
increasing $R_{\rm dis}/R_c$. 

We can compare our results with those obtained in 
Refs.~\cite{SSKK-02,PVCL-08}. By using 
density-functional methods, Ref.~\cite{SSKK-02} studied the 
model with $R_{\rm dis}/R_c = 1$ at the slightly
lower value of $f$,  $f = 0.37$ (corresponding to $\eta_{\rm dis} = 0.05$),
and several values of $q$, $q = 0.3$, 0.6, 1.0 
(none of them agrees unfortunately with ours). 
Their results are not consistent with ours. They find that 
in all cases the binodal curve in the presence of the matrix is below that
in the bulk, while here we find the opposite except deep in the 
colloid-liquid region, i.e. for $\eta_c \gtrsim 0.20$. 
Moreover, they find that  $\eta_{c,\rm crit}$ decreases with increasing $f$
in all cases, which is again in contrast with our results.
On the other hand we are fully consistent with the 
results of Ref.~\cite{PVCL-08}, which study the model for 
$R_{\rm dis}/R_c = 1$, $f = 0.37$ ($\eta_{\rm dis} = 0.05$), and 
$q = 0.8$. For the critical point they obtain
$\eta_{c,\rm crit} = 0.192$ and $\eta_{p,\rm crit} = 0.292$, confirming
that $\eta_{c,\rm crit}$ increases in the presence of disorder. 
From a quantitative point of view, their critical-point estimates are 
fully consistent with ours. In particular, the naive extrapolation 
we have performed above apparently
estimates correctly the critical-point position at the 5\% level.

\section{Conclusions} \label{sec4}

In this paper we determine the fluid-fluid demixing curves 
for the AO model in a colloidal matrix for $q=0.8$ and 
several values of $R_{\rm dis}/R_c$ and $f$ (equivalently, of the 
disorder concentration $c_{\rm dis}$). This study should provide quantitative
informations on the phase behavior 
of a polymer-colloid mixture in a porous material close to the 
$\theta$ point. Our main results are the following: 
\begin{itemize}
\item Disorder is specified by two parameters: the disorder packing fraction 
$\eta_{\rm dis}$ and the ratio $R_{\rm dis}/R_c$. 
At least for $R_{\rm dis}/R_c \le 1$, the parameter range we consider, 
most of the disorder dependence of the results can be parametrized by 
using a single parameter, the effective matrix filled-space ratio $f$, 
which gives the volume fraction unavailable to the colloids.
In the $z_p$, $z_c$ 
plane (or equivalently in terms of the reservoir packing fractions 
$\eta_c^r$ and $\eta_p^r$) the coexistence curve depends essentially only
on $f$.
\item It is possible to observe capillary condensation of the colloids.
For certain values of the parameters a colloid-gas bulk phase is in
equilibrium with a colloid-liquid phase in the matrix.
\item At least for $f \lesssim 0.4$ the binodal curves expressed in 
terms of the packing fractions (system representation) 
show a relatively small dependence on disorder.
The critical point instead changes significantly. The critical colloid packing 
fraction $\eta_{c,\rm crit}$ is, to a large extent, only a function of $f$
and it increases as $f$ increases. 
The critical polymer packing
fraction $\eta_{p,\rm crit}$ depends instead both on $f$ and $R_{\rm dis}/R_c$.
At fixed $f$ it decreases as $R_{\rm dis}/R_c$ increases.
\end{itemize}

\bigskip

The authors gratefully acknowledge extensive discussions with Ettore Vicari.
The MC simulations were performed at the INFN Pisa GRID DATA center and on 
the INFN cluster CSN4.

\appendix
\section{Monte Carlo simulations: some technical details} \label{app}

We have performed simulations in the grand-canonical (GC) ensemble,
which physically describes a system adsorbed in a colloid matrix
in chemical equilibrium with a reservoir of pure polymers
and a reservoir of pure noninteracting colloids.
The basic parameters are the colloid and polymer fugacities 
$z_c$ and $z_p$. In the bulk the partition function is
\begin{equation}
\Xi(V,z_p,z_c) = \sum_{N_p,N_c} z_p^{N_p} z_c^{N_c} Q(V,N_p,N_c),
\label{GCdist}
\end{equation}
where $Q(V,N_p,N_c)$ is the configurational partition function of 
a system of $N_p$ polymers and $N_c$ colloids in a volume $V$. 
We drop the irrelevant thermal length so that $Q(V,1,0) = Q(V,0,1) = V$.
In the presence of first-order transitions it is quite difficult 
to sample correctly the GC distribution. To bypass the 
difficulties  we use the umbrella-sampling (sometimes also called 
multicanonical) method \cite{TV-77}. Instead of generating configurations
with the GC weight, we use an umbrella distribution
\begin{equation}
{1\over \pi(N_c)} z_p^{N_p} z_c^{N_c} e^{-\beta H},
\label{umbrella-dist}
\end{equation}
with a properly chosen $\pi(N_c)$ which is defined below. If 
$\langle \cdot\rangle_{GC}$ and $\langle \cdot\rangle_\pi$ are the 
averages with respect to the GC distribution and to the 
distribution (\ref{umbrella-dist}), respectively, we have
\begin{equation}
\langle O(N_c,N_p) \rangle_{GC} = 
  {\langle \pi(N_c) O(N_c,N_p) \rangle_\pi \over \langle \pi(N_c) \rangle_\pi}
.
\end{equation}
This relation allows us to obtain GC averages from simulations
using the distribution (\ref{umbrella-dist}). The function $\pi(N_c)$ must
be chosen so that in the simulation 
the system can move easily between the two phases.
Consider the histogram of $N_c$ in the GC distribution,
i.e.
\begin{equation} 
  h(N_{c,0}) = \langle \delta(N_c,N_{c,0})\rangle_{GC},
\end{equation}
where $\delta(x,y)$ is Kronecker's delta. Assume that the 
system is close to phase separation so that $h(N_{c})$ has two peaks at
$N_{c,\rm min}$ (colloid-gas phase) and 
$N_{c,\rm max}$ (colloid-liquid phase).
The optimal choice is then
\begin{equation}
\begin{array}{ll}
  \pi(N) = a h(N_{c,\rm min})   & \qquad N \le N_{c,\rm min}, \\
  \pi(N) = a h(N)            & \qquad N_{c,\rm min} \le N \le N_{c,\rm max}, \\
  \pi(N) = a h(N_{c,\rm max})   & \qquad N \ge N_{c,\rm max} ,
\end{array}
\end{equation}
where $a$ is an irrelevant constant.
Indeed, if $N_{c,\rm min} \le N_c \le N_{c,\rm max}$, the observed 
histogram in the umbrella distribution is flat, i.e. independent of $N_c$.
Hence, the system can move freely between the two phases, allowing 
a precise determination of any required thermodynamic property.

In order to determine the colloid fugacity $z_c^*$ at
coexistence for a given value of the polymer fugacity $z_p$,
we consider $N_m$ colloid fugacities $\{z_{c,m}\}$, such that 
for $z_{c,1}$ ($z_{c,N_m}$)
the system is in the colloid-gas (colloid-liquid) phase. 
Then, we determine the umbrella functions $\pi_m(N_c)$ iteratively.
First, we perform a short hysteresis cycle
in which we perform $N_{\rm therm}$ 
GC iterations at $z_c = z_{c,1}$, then at $z_{c,2}$, and so on,
 up to $z_{c,N_m}$; then we decrease $z_{c}$ till we reach
again $z_{c,1}$.  If $h_m^{(1),+}(N_c)$ and $h_m^{(1),-}(N_c)$
are the histograms obtained 
at $z = z_m$ (the + refers to the distribution obtained while 
increasing $z_c$ and the $-$ to that obtained while decreasing the 
fugacity), we set $h_m^{(1)}(N_c) = h_m^{(1),+}(N_c) + h_m^{(1),-}(N_c)$ and
\begin{equation}
\begin{array}{ll}
   \pi_m^{(1)}(N_c) = h_m^{(1)}(N_c)/M  & 
    \qquad \hbox{if}\,\, h_m^{(1)}(N_c) \ge M,  \\
   \pi_m^{(1)}(N_c) =  1 &  
    \qquad \hbox{if}\,\, h_m^{(1)}(N_c) \le M,  
\end{array}
\end{equation}
where $M \equiv \max_{N_c}[h_m^{(1)}(N_c)]/10$. 
Then, we repeat again the same hysteresis cycle several times. 
At iteration $k$, for each $z_{c,m}$ we perform the simulation using 
the distribution (\ref{umbrella-dist}) with $\pi = \pi_m^{(k-1)}$.
Then, we set
\begin{equation}
\begin{array}{ll}
   \pi_m^{(k)}(N_c) = \pi_m^{(k-1)}(N_c) h_m^{(k)}(N_c)/M  & 
    \qquad \hbox{if}\,\, h_m^{(k)}(N_c) \ge M,  \\
   \pi_m^{(k)}(N_c) = \pi_m^{(k-1)}(N_c) & 
    \qquad \hbox{if}\,\, h_m^{(k)}(N_c) \le M,  
\end{array}
\end{equation}
where $M \equiv \max_{N_c}[h_m^{(k)}(N_c)]/10$. We stop when we observe that,
for at least some values of $m$, $h_m^{(k),+}(N_c)$ and $h_m^{(k),-}(N_c)$
are nonvanishing in an interval of values of $N_c$ that extends 
between the two phases.

Once we have a reasonable estimate of the functions $\pi_m(N_c)$,
we could just perform an extensive simulation at single value of $z_{c,m}$, 
(an optimal choice would 
be to take the value for which $\pi_m(N_c)$ is clearly bimodal).
Data for different values of $z_c$ could just be obtained by 
standard reweighting techniques. However, we have found more convenient,
to use all information we have collected and simulate 
all systems together, using the simulated-tempering method \cite{MP-92}. 
Note that, in the standard implementation of the method, one 
should be careful that the fugacities $z_{c,m}$ are such that the 
colloid-number distributions overlap; otherwise, no fugacity swap
is accepted. In our case, since we use umbrella distributions, 
the overlap condition is always verified, and thus the number $N_m$ of 
needed systems is always small. Typically we take $N_m = 10$. 
If 
\begin{equation}
\Xi_{\pi_m}(V,z_p,z_{c,m}) = 
\sum_{N_p,N_c} {z_p^{N_p} z_{c,m}^{N_c}\over \pi_m(N_c)} Q(V,N_p,N_c),
\end{equation}
we consider the extended partition function 
\begin{equation}
\Xi^{ST} = \sum_m f_m \Xi_{\pi_m}(V,z_p,z_{c,m}).
\end{equation}
The constants $f_m$ are chosen so that all terms in the sum are approximately
equal. If we require 
\begin{equation}
f_m \Xi_{\pi_m}(V,z_p,z_{c,m}) = f_{m-1} \Xi_{\pi_{m-1}}(V,z_p,z_{c,m-1}),
\end{equation}
we obtain 
\begin{equation}
{f_m\over f_{m-1}} = R_m \qquad 
{f_{m-1}\over f_{m}} = S_m ,
\end{equation}
with
\begin{eqnarray}
R_m &\equiv& \left\langle 
    \left( {z_{c,m-1}\over z_{c,m}}\right)^{N_c}
           {\pi_m(N_c) \over \pi_{m-1}(N_c)} 
         \right\rangle_{\pi,m}, \\
S_m &\equiv& \left\langle \left( {z_{c,m}\over z_{c,m-1}}\right)^{N_c}
           {\pi_{m-1}(N_c) \over \pi_m(N_c)}
         \right\rangle_{\pi,m-1},
\end{eqnarray}
where $\langle\cdot\rangle_{\pi,m}$ indicates the mean value with respect to 
the umbrella distribution (\ref{umbrella-dist}) with $z_c = z_{c,m}$,
$\pi = \pi_m$. 
Combining these expressions we define the ratios as
\begin{equation}
{f_m\over f_{m-1}} =  \sqrt{R_m/S_m}.
\label{fm-fm-1}
\end{equation}
The constants $R_m$ and $S_m$ are determined together with the 
umbrella sampling functions $\pi_m$. Then, we set 
$f_1 = 1$ and use Eq.~(\ref{fm-fm-1}) to determine the constants $f_m$,
$m\ge 2$.

In the matrix case, the GC partition function is still given
be Eq.~(\ref{GCdist}), with the only difference that 
one should take into account the interactions between the freely
moving particles and the matrix. Since the GC partition function
depends on the matrix, also the functions $\pi_m$ and the 
constants $f_m$ are matrix dependent. Thus, 
we recompute them when we restart the simulation 
with the different matrix.

In the MC simulations we take $N_m \approx 10$. One 
MC iteration consists in 3 fugacity swaps and 1000-5000 
GC moves in which colloids and polymers are inserted or 
removed. For this purpose we use the cluster move discussed in 
Ref.~\cite{Vink-04} together with standard moves in which a single
polymer is removed or inserted. 
For each disorder instance, we perform $N_{\rm ini}$ iterations 
to determine the umbrella functions and then
$N_{\rm iter}$ iterations to measure several histograms. 
Typically, $N_{\rm ini}$ varies between $5000 N_m$ and 
$20000 N_m$, while  $N_{\rm iter}$ is of the order of 
$20000 N_m$.

In the simulation we determine the colloid and polymer histograms for 
a large number (typically 100) of colloid fugacities $z_{c,r}$. 
They are obtained by measuring, 
for each matrix realization $\alpha$, the 
reweighted histograms $p_c(\alpha,z_{c,r},N_{c,0})$ and 
$p_p(\alpha,z_{c,r},N_{p,0})$:
\begin{eqnarray}
p_c(\alpha,z_{c,r},z_{c,m},N_{c,0}) &=&
   \sum_i \left({z_{c,r}\over z_{c,m}}\right)^{N_{c,i}} 
     \pi_m(N_{c,i}) \delta (N_{c,i},N_{c,0}) \delta(z_{c,m},z_{c,i}) ,
\\
p_p(\alpha,z_{c,r},z_{c,m},N_{p,0}) &=& 
   \sum_i \left({z_{c,r}\over z_{c,m}}\right)^{N_{c,i}} 
     \pi_m(N_{c,i}) \delta (N_{p,i},N_{p,0}) \delta(z_{c,m},z_{c,i}) ,
\end{eqnarray}
where $i$ refers to the MC iteration, and 
$N_{p,i}$, $N_{c,i}$, $z_{c,i}$ are the number of polymers and  colloids 
and the colloid fugacity at the $i$th iteration.
The colloid histogram is then
\begin{equation}
h_{c,\rm ave}(N_{c,0},z_p,z_{c,r}) = 
{1\over N_{\alpha}} \sum_\alpha 
\left[ {p_c(\alpha,z_{c,r},z_{c,m},N_{c,0}) \over 
       \sum_{N_c} p_c(\alpha,z_{c,r},z_{c,m},N_{c}) } \right],
\label{estima-1}
\end{equation}
where $N_\alpha$ is the number of matrix realizations. 
Note that we obtain a different estimate of the distributions at 
$z_{c,r}$ for each of the $z_{c,m}$. 

As a final comment, note that our estimates (\ref{estima-1})
are biased, since they are disorder averages of a ratio of thermal averages. 
This means that, 
if we take the limit $N_\alpha \to\infty$ at fixed $N_{\rm iter}$, we obtain
estimates that differ from the correct result by a term (the {\em bias}) of 
order $1/N_{\rm iter}$. One could perform a bias correction, as discussed in
Ref.~\cite{HPPV-07}. However, given the small number of disorder instances,
we have found that in the present case the bias correction is not relevant.


\begin{thebibliography}{199}


\bibitem{Poon-02} 
W. C. K. Poon, 
J. Phys.: Condensed Matter {\bf 14}, R859 (2002).

\bibitem{FS-02}
M. Fuchs and K. S. Schweizer, 
J. Phys.: Condensed Matter {\bf 14}, R239 (2002).

\bibitem{TRK-03}
R. Tuinier, J. Rieger, and C. G. de Kruif,
Adv. Colloid Interface Sci. {\bf 103}, 1 (2003).

\bibitem{MvDE-07}
K. J. Mutch, J. S. van Duijneveldt, and J. Eastoe,
Soft Matter {\bf 3}, 155 (2007).

\bibitem{FT-08}
G. J. Fleer and R. Tuinier,
Adv. Coll. Interface Sci. {\bf 143}, 1 (2008).

\bibitem{ME-09}
O. Myakonkaya and J. Eastoe,
Adv. Coll. Interface Sci. {\bf 149}, 39 (2009).


\bibitem{LLPR-03}
T. C. Lee, J. T. Lee, D. R. Pilaski, and M. Robert,
Physica A {\bf 329}, 411 (2003).

\bibitem{VvDV-03}
G. A. Vliegenthart, J. S. van Duijneveldt, and B. Vincent,
Faraday Discuss. {\bf 123}, 65 (2003).

\bibitem{SCSZ-03}
S. A. Shah, Y. L. Chen, K. S. Schweizer, and C. F. Zukoski,
J. Chem. Phys. {\bf 118}, 3350 (2003).

\bibitem{KSMH-04}
T. Kramer, S. Scholz, M. Maskros, and K. Huber,
J. Colloid Interface Sci. {\bf 279}, 447 (2004).

\bibitem{LCP-04}
I. Lynch, S. Cornen, and L. Piculell, 
J. Phys. Chem. B {\bf 108}, 5443 (2004).

\bibitem{HEAD-05}
Y. Hennequin, M. Evens, C. M. Q. Angulo, and J. S. van Duijneveldt,
J. Chem. Phys. {\bf 123}, 054906 (2005).

\bibitem{KSH-05}
T. Kramer, R. Schweins, and K. Huber,
J. Chem. Phys. {\bf 123}, 014903 (2005);
Macromolecules {\bf 38}, 151 (2005); 
{\bf 38}, 9783 (2005).

\bibitem{ZvD-06}
Z. Zhang and J. S. van Duijneveldt,
Langmuir {\bf 22}, 63 (2006).

\bibitem{LPKCSBFE-09}
M. Laurati, G. Petekidis, N. Koumakis, F. Cardineaux,
A. B. Schofield, J. M. Brader, M. Fuchs, and S. U. Egelhaaf,
J. Chem. Phys. {\bf 130}, 134907 (2009).

\bibitem{MvDEGH-09-10}
K. J. Mutch, J. S. van Duijneveldt, J. Eastoe, I. Grillo, and R. K. Heenan,
Langmuir {\bf 25} 3944 (2009); {\bf 26}, 1630 (2010).


\bibitem{GHR-83}
A. P. Gast, C. K. Hall, and W. B. Russell, 
J. Colloid Interface Sci. {\bf 96}, 251 (1983).

\bibitem{LPPSW-92}
H. N. W. Lekkerkerker, W. C. K. Poon, P. N. Pusey, 
A. Stroobants, and P. B. Warren, 
Europhys. Lett. {\bf 20}, 559 (1992).

\bibitem{MF-94}
E. J. Meijer and D. Frenkel, J. Chem. Phys. {\bf 100}, 6873 (1994).

\bibitem{Sear-97-02}
R. P. Sear, Phys. Rev. E {\bf 56}, 4463 (1997);
Phys. Rev. E {\bf 66}, 051401 (2002).

\bibitem{DBE-99}
M. Dijkstra, J. M. Brader, and R. Evans, 
J. Phys.: Condensed Matter {\bf 11}, 10079 (1999).

\bibitem{FS-00}
M. Fuchs and K. S. Schweizer,
Europhys. Lett. {\bf 51}, 621 (2000).

\bibitem{SLBE-00}
M. Schmidt, H. L\"owen, J. M. Brader, and R. Evans,
Phys. Rev. Lett. {\bf 85}, 1934 (2000);
J. Phys.: Condensed Matter {\bf 14}, 9353 (2002).

\bibitem{BLH-02}
P. G. Bolhuis, A. A. Louis, and J. P. Hansen, 
Phys. Rev. Lett. {\bf 89}, 128302 (2002).

\bibitem{DvR-02}
M. Dijkstra and R. van Roij, 
Phys. Rev. Lett. {\bf 89}, 208303 (2002). 

\bibitem{DLL-02}
J. Dzubiella, C. N. Likos, and H. L\"owen, 
J. Chem. Phys. {\bf 116}, 9518 (2002). 

\bibitem{BLM-03}
P. G. Bolhuis, A. A. Louis, and E. J. Meijer,
Phys. Rev. Lett. 90, 068304 (2003).  

\bibitem{MJLBR-03}   
A. Moncho-Jorda, A. A. Louis, P. G. Bolhuis, and R. Roth
J. Phys.: Condensed Matter {\bf 48}, S3429 (2003).

\bibitem{Tuinier-03}
R. Tuinier, Eur. Phys. J. E {\bf 10}, 123 (2003).

\bibitem{PVJ-03} 
P. Paricaud, S. Varga, and G. Jackson,
J. Chem. Phys. {\bf 118}, 8525 (2003).

\bibitem{VH-04bis}
R. L. C. Vink and J. Horbach, 
J. Chem. Phys. {\bf 121}, 3253 (2004).

\bibitem{VHB-05}
R. L. C. Vink, J. Horbach, and K. Binder,
Phys. Rev. E {\bf 71}, 011401 (2005).

\bibitem{VS-05}
R. L. C. Vink and M. Schmidt, 
Phys. Rev. E 71, 051406 (2005).

\bibitem{VJDL-05}
R. L. C. Vink, A. Jusufi, J. Dzubiella, and C. N. Likos,
Phys. Rev. E 72, 030401(R) (2005).

\bibitem{Bryk-05}
P. Bryk, J. Chem. Phys. {\bf 122}, 064902 (2005).

\bibitem{FS-05}
M. Fasolo and P. Sollich,
J. Phys.: Condensed Matter {\bf 17}, 797 (2005).

\bibitem{PH-05}
A. Pelissetto and J. P. Hansen,
Macromolecules {\bf 39}, 9571 (2006).

\bibitem{CVPR-06}
C.-Y. Chou, T. T. M. Vo, A. Z. Panagiotopoulos, and M. Robert,
Physica A {\bf 369}, 275 (2006).

\bibitem{FT-07}
G. J. Fleer and R. Tuinier, 
Phys. Rev. E {\bf 76}, 041802 (2007).

\bibitem{ZVBHV-09}
J. Zausch, P. Virnau, K. Binder, J. Horbach, R. L. C. Vink,
J. Chem. Phys. {\bf 130}, 064906 (2009).


\bibitem{AO-54}
S. Asakura and F. Oosawa, J. Chem. Phys. {\bf 22}, 1255 (1954).

\bibitem{Vrij-76}
A. Vrij, Pure and Appl. Chem. {\bf 48}, 471 (1976).

\bibitem{footnote-theta} 
This is correct only for infinite-length polymers. For finite-length 
chains, polymers interact weakly (as an inverse power of $\ln L$, where 
$L$ is the degree of polymerization) and 
a proper coarse-grained description requires the introduction of an attractive 
pair potential and of a repulsive (needed for thermodynamic stability)
three-body potential, see 
V. Krakoviack, J. P. Hansen, and A. A. Louis, 
Phys. Rev. E {\bf 67}, 041801 (2003);
C. I. Addison, A. A. Louis, and J. P. Hansen, 
J. Chem. Phys. {\bf 121}, 612 (2004);
A. Pelissetto and J. P. Hansen,
J. Chem. Phys. {\bf 122}, 134904 (2005).

\bibitem{footnote-expt-porousmat} For a list of experimental
studies of binary mixtures in porous materials, see the 
references cited in 
E. Sch\"oll-Paschinger, D. Levesque, J.-J. Weis, and G. Kahl,
Phys. Rev. E {\bf 64}, 011502 (2001).


\bibitem{SSKK-02}
M. Schmidt, E. Sch\"oll-Paschinger, J. K\"ofinger, and G. Kahl,
J. Phys.: Condensed Matter {\bf 14}, 12099 (2002).

\bibitem{VBL-06}
R. L. C. Vink, K. Binder, and H. L\"owen,
Phys. Rev. Lett. {\bf 97}, 230603 (2006).

\bibitem{VBL-08}
R. L. C. Vink, K. Binder, and H. L\"owen,
J. Phys.: Condensed Matter {\bf 20}, 404222 (2008).

\bibitem{PVCL-08}
G. Pellicane, R. L. C. Vink, C. Caccamo, and H. L\"owen, 
J. Phys.: Condensed Matter {\bf 20}, 115101 (2008).


\bibitem{Vink-09}
R. L. C. Vink, Soft Matter {\bf 5}, 4388 (2009).

\bibitem{deGennes-84}
P. G. de Gennes, J. Phys. Chem. {\bf 88}, 6469 (1984).

\bibitem{DSLP-08}
P. G. De Sanctis Lucentini and G. Pellicane,
Phys. Rev. Lett. {\bf 101}, 246101 (2008).

\bibitem{FV-11}
T. Fischer and R. L. C. Vink,
J. Chem. Phys. {\bf 134}, 055106 (2011).


\bibitem{SZ-70}
H. Scher and R. Zallen,
J. Chem. Phys. {\bf 53}, 3759 (1970).


\bibitem{HR-67}
W. G. Hoover and F. H. Ree, 
J. Chem. Phys. {\bf 47}, 4873 (1967).

\bibitem{TV-77}
G. M. Torrie and J. P Valleau,
J. Comp. Phys. {\bf 23}, 197 (1977).


\bibitem{MP-92}
E. Marinari and G. Parisi, Europhys. Lett. {\bf 19}, 451 (1992).



\bibitem{Vink-04}
R. L. C. Vink in 
``Computer simulation studies in condensed matter physics XVIII,"
edited by D. P. Landau, S. P. Lewis, and H. B. Schuettler 
(Springer, Berlin, 2004).

\bibitem{footnote-etacr} 
We use the Carnahan-Starling expression to relate the colloid 
reservoir packing fraction $\eta_c^r$ to the fugacity $z_c$: 
$z_c R_c^3 = 3 \eta_c^3/(4 \pi) \exp[f(\eta_c^r)]$ with 
$f(\eta) = \eta (8 - 9\eta + 3\eta^2)/(1 - \eta)^3$; 
see N. F. Carnahan and K. E. Starling, J. Chem. Phys. {\bf 51}, 635 (1969);
L. L. Lee,  J. Chem. Phys. {\bf 103}, 9388 (1995).

\bibitem{VTB-10}
R. L. C. Vink, T. Fischer, and K. Binder, 
Phys. Rev. E {\bf 82}, 051134 (2010).

\bibitem{FV-11b}
T. Fischer and R. L. C. Vink, 
J. Phys.: Condensed Matter {\bf 23}, 234117 (2011).

\bibitem{HPPV-07}
M. Hasenbusch, F. Parisen Toldin, A. Pelissetto, and E. Vicari,
J. Stat. Mech.: Theory Exp. P02016 (2007).

\end{thebibliography}
\end{document}